\documentclass[sigconf]{acmart}

\usepackage[english]{babel}
\usepackage{blindtext}

\usepackage{subfigure,algorithm,algorithmic}

\usepackage{bbm}
\usepackage{balance}
\usepackage[font=small,labelfont=bf]{caption}

\usepackage{color,xcolor,caption}
\usepackage{color,soul}
\usepackage{subfigure}
\usepackage{balance}
\usepackage{graphicx}
\usepackage{url}
\usepackage{multirow}
\usepackage{graphicx}
\usepackage{amsfonts}
\usepackage{caption}
\usepackage{wrapfig}
\usepackage{alltt}
\usepackage{booktabs}
\usepackage{xspace}
\usepackage{verbatim}
\usepackage{footnote}
\usepackage{epstopdf}
\usepackage{float,fancyvrb}
\usepackage{enumitem}

\usepackage{pifont}
\newcommand{\cmark}{\ding{51}}%
\newcommand{\xmark}{\ding{55}}%

\newcommand{\presec}{\vspace{0.1in}}
\newcommand{\postsec}{\vspace{0.1in}}
\newcommand{\presub}{\vspace{0.05in}}
\newcommand{\postsub}{\vspace{0.05in}}
\newcommand{\presubsub}{\vspace{0.00in}}
\newcommand{\postsubsub}{\vspace{0.00in}}
\newcommand{\precaption}{\vspace{0.1in}}
\newcommand{\postcaption}{\vspace{0.1in}}

\providecommand{\ie}{\emph{i.e.,} }
\providecommand{\eg}{\emph{e.g.,} }

\providecommand{\etal}{\emph{et al.}\xspace}

\newcommand{\googlePlay}{{Google Play Store}\xspace}
\newcommand{\iosstore}{{Apple App Store}\xspace}
\newcommand{\honeyapp}{{honey app}\xspace}

\providecommand{\offertoro}{OfferToro\xspace}
\providecommand{\ayet}{ayeT-Studios\xspace}
\providecommand{\adscend}{AdscendMedia\xspace}
\providecommand{\hangmyads}{HangMyAds\xspace}
\providecommand{\rankapp}{RankApp\xspace}
\providecommand{\adgem}{AdGem\xspace}
\providecommand{\fyber}{Fyber\xspace}

\newcommand{\installonly}{{\textit{no activity}}\xspace}
\newcommand{\register}{\textit{{registration}}\xspace}
\newcommand{\appusage}{\textit{{usage}}\xspace}

\newcommand{\apppurchase}{\textit{{purchase}}\xspace}

\newcommand{\activity}{{\textit{activity}}\xspace}
\newcommand{\Activity}{{\textit{Activity}}\xspace}

\newcommand{\Installonly}{{\textit{No activity}}\xspace}
\newcommand{\Register}{{\textit{Registration}}\xspace}
\newcommand{\Appusage}{\textit{{Usage}}\xspace}
\newcommand{\Apppurchase}{\textit{{Purchase}}\xspace}

\newcommand{\incentservice}{{IIP}\xspace}
\newcommand{\incentservices}{{IIPs}\xspace}

\newcommand{\shehroze}[1]
{{\color{black} #1}}

\providecommand{\ie}{\emph{i.e.,} }
\providecommand{\eg}{\emph{e.g.,} }

\providecommand{\etal}{\emph{et al.}\xspace}

\copyrightyear{2020}
\acmYear{2020}
\setcopyright{acmcopyright}\acmConference[IMC '20]{ACM Internet Measurement Conference}{October 27--29, 2020}{Virtual Event, USA}
\acmBooktitle{ACM Internet Measurement Conference (IMC '20), October 27--29, 2020, Virtual Event, USA}
\acmPrice{15.00}
\acmDOI{10.1145/3419394.3423662}
\acmISBN{978-1-4503-8138-3/20/10}


\begin{document}

\author{Shehroze Farooqi}
\affiliation{%
  \institution{University of Iowa/ICSI}
}
\author{Álvaro Feal}
\affiliation{%
  \institution{IMDEA Networks Institute / Universidad Carlos III de Madrid}
}
\author{Tobias Lauinger}
\affiliation{%
  \institution{NYU}
}
\author{Damon McCoy}
\affiliation{%
  \institution{NYU}
}
\author{Zubair Shafiq}
\affiliation{%
  \institution{University of California, Davis}
}
\author{Narseo Vallina-Rodriguez}
\affiliation{%
  \institution{IMDEA Networks Institute/ICSI}
}

\title{Understanding Incentivized  Mobile App Installs \\on Google Play Store}

\title[Understanding Incentivized  Mobile App Installs]{Understanding Incentivized  Mobile App Installs \\on Google Play Store} 
\renewcommand{\shortauthors}{Shehroze Farooqi et al.}

\begin{abstract}

``Incentivized'' advertising platforms allow mobile app developers to acquire new users by directly paying users to install and engage with mobile apps (\eg create an account, make in-app purchases).
Incentivized installs are banned by the Apple App Store and discouraged by the Google Play Store because they can manipulate app store metrics (\eg install counts, appearance in top charts). Yet, many organizations still offer incentivized install services for Android apps. 
In this paper, we present the first study to understand the ecosystem of incentivized mobile app install campaigns in Android and its broader ramifications through a series of measurements. 
We identify incentivized install campaigns that require users to install an app and perform in-app tasks targeting manipulation of a wide variety of user engagement metrics (\eg daily active users, user session lengths) and revenue.
Our results suggest that these artificially inflated metrics can be effective in improving app store metrics as well as helping mobile app developers to attract funding from venture capitalists.
Our study also indicates lax enforcement of the Google Play Store's existing policies to prevent these behaviors. It further motivates the need for stricter policing of incentivized install campaigns. Our proposed measurements can also be leveraged by the Google Play Store to identify potential policy violations.

\end{abstract}

\maketitle

\presec \section{Introduction} \postsec

Popular app stores such as the \googlePlay and the \iosstore list over 2 million mobile apps \cite{play_billion,apple_ios_users}.
The proliferation of mobile apps has increased the competition among app developers to improve the ``visibility'' of their apps in app stores' searches and top charts to acquire new users. Creating and retaining a solid user base is critical for maximizing the revenue of mobile apps, mostly through in-app advertising or in-app purchases  \cite{iosBusinessModel,gplayBusinessModel}.\footnote{\googlePlay and \iosstore had an estimated revenue of 83 billion dollars in 2019 \cite{appstoresRevenuesDigitalInfo}.}
Furthermore, popular mobile apps with strong user engagement metrics (\eg install counts, daily active users) and revenue will also be in a better position to attract funding from venture capitalists (VCs) \cite{investorUser,investorGrowthForbes,investorInstallsStartups}.

Developers spent more than 14 billion dollars in 2019 on various types of advertising campaigns for promoting mobile app installs \cite{appsFlyerAdSpent}. Oftentimes, this is done through traditional advertising-based models~\cite{appAdPlatforms}. 
Another type of advertising is ``incentivized'' mobile app install campaigns, where users install an advertised mobile app and perform in-app tasks in exchange for monetary (\eg gift card, PayPal balance) or non-monetary (\eg in-app points or virtual currency) rewards \cite{incentivize_discussion}.
Mobile app developers pay incentivized install platforms (or \incentservices) to advertise an incentivized install offer, which is distributed to end users through a network of ``affiliate'' apps.
These incentivized install campaigns are orders of magnitude cheaper in comparison to regular, non-incentivized mobile app install campaigns --- on average, a mobile app install through incentivized advertising costs around \$0.06 (as we show later in the paper), as opposed to \$1.22 for non-incentivized installations \cite{appInstallCost}.

The incentivized install ecosystem is highly sophisticated and opaque. Incentivized installs have a poor reputation since users (potentially ``crowd workers'') are likely performing actions for a monetary reward rather than based on a genuine interest in an app. Furthermore, the ability to allocate specific in-app tasks to users (\eg creating an account, completing a level of a game, performing in-app purchases) through incentivized install campaigns enables efficient manipulation of targeted user engagement metrics (\eg number of active users, user session lengths) and revenues. These inflated user engagement metrics can help apps manipulate app store metrics (install counts, visibility), which might lead to potential violations of app store policies \cite{incentInstallPolicyGoogle,iOSappStorePolicy}. 
In fact, app developers might abuse these schemes to deceitfully monetize through in-app advertising or obtain VC funding based on inflated user engagement or revenue metrics. 
It is noteworthy that incentivized installs are banned by the \iosstore and discouraged by the \googlePlay.
Note that the \googlePlay allows their use as long as they are not used to manipulate app store metrics (\eg install counts, visibility) \cite{incentInstallsrankingsGoogleBlog}. As a result, \incentservices often do not support the \iosstore, but potentially generate millions of USD in revenue on the \googlePlay \cite{fyberOwler,fyberReport}.

It can be challenging for app stores to detect incentivized installs since these installs and user actions resemble that of authentic organic users~\cite{engagementVisibilityGplay_2}. Prior work \cite{zhu2013ranking,chen17collusiveRankingASIACCS,rahman17fairPlaySIAM,rahman18searchfraudHT,DouDownFraud19,rahman19artCraftCCS} has studied manipulation of app store metrics on the \googlePlay, focusing primarily on detecting installs driven by automation or by low-cost labor recruited outside of the app store. 
However, to the best of our knowledge, there has been no systematic investigation of incentivized installs. As a result, we lack a good understanding of the incentivized install ecosystem and the effectiveness of these campaigns to manipulate user engagement metrics. More importantly, it is unclear whether or not existing app store defenses are able to detect the manipulation of app store metrics (\eg install counts, visibility). 

To fill this gap, this paper studies the ecosystem of incentivized installs, its prevalence, and its potential impact.  We narrow our focus on studying the ecosystem of incentivized installs offering monetary rewards on the \googlePlay. We measure the effectiveness of these campaigns and shed light on app stores' existing defenses \cite{incentInstallsrankingsGoogleBlog,spamInstallsGoogleBlog} to detect incentivized installs.
First, we perform measurements by purchasing incentivized installs from multiple \incentservices for our purpose-built \honeyapp that we publish on the \googlePlay. 
This allows us to understand the behavior and effectiveness of incentivized installs through the lens of mobile app developers and to evaluate existing enforcement by \googlePlay. Then, we perform longitudinal measurements to monitor incentivized install campaigns of mobile apps on \incentservices over a period of three months, at scale. This complementary perspective allows us to characterize incentivized install campaigns of advertised mobile apps, and to understand the potential benefits of running these campaigns.  

The main contributions of our study are the following:

\begin{enumerate}[leftmargin=*]

    \item Our measurements suggest that most of the users are crowd workers installing apps advertised on \incentservices to earn money. 
    The results suggest a lack of enforcement from the \googlePlay to detect these practices since our purchased installs increased the total install count of our \honeyapp on the \googlePlay from 0 to over 1,000. 

    \item Through longitudinal measurements, we identify campaigns of 922 apps that advertised a variety of incentivized install offers. We find that more than half of these offers require users to perform in-app tasks targeting manipulation of a wide variety of user engagement metrics (\eg install counts, increase user session lengths) and revenues. These behaviors seem to contradict Google Play Store's policies.

    \item Our results show that there is a correlation between apps advertised on \incentservices and the improvement of app store metrics on the \googlePlay.  In comparison to a baseline dataset of regular Android apps, apps advertised on \incentservices show up to 8$\times$ increase in install counts and appear in top charts up to 2$\times$ more frequently. Our results also demonstrate that there is a correlation between advertising apps on \incentservices and developers of advertised apps raising funding: apps advertised on \incentservices tend to raise funding up to 2$\times$ more frequently.

    \item Finally, we find that artificial in-app activity generated through incentivized install campaigns provides opportunities to developers of mobile apps to artificially increase gross revenue through arbitrage and advertising. For example, more than 60\% of apps requiring users to perform in-app tasks integrate 5 or more advertising libraries. 

\end{enumerate}

Our results show that \incentservices can have negative effects on the app store and beyond (\eg deceitfully raise funding and erode consumer trust) due to their ability to manipulate user engagement metrics and revenue. In fact, some \incentservices even openly advertise services that seem to violate \googlePlay's policies. Our study motivates the need for stricter policies against \incentservices, similar to those implemented in the \iosstore \cite{fyberiOSIncentInstallBan,appleRejectAppsPocktgmr}.
Our measurement methodology to monitor \incentservices can also be leveraged by the \googlePlay to identify potential policy violations.


\section{Incentivized Installs} \label{sec: incentecosystem}

\begin{figure*}[!t]
    \centering
    \includegraphics[width=2.0\columnwidth]{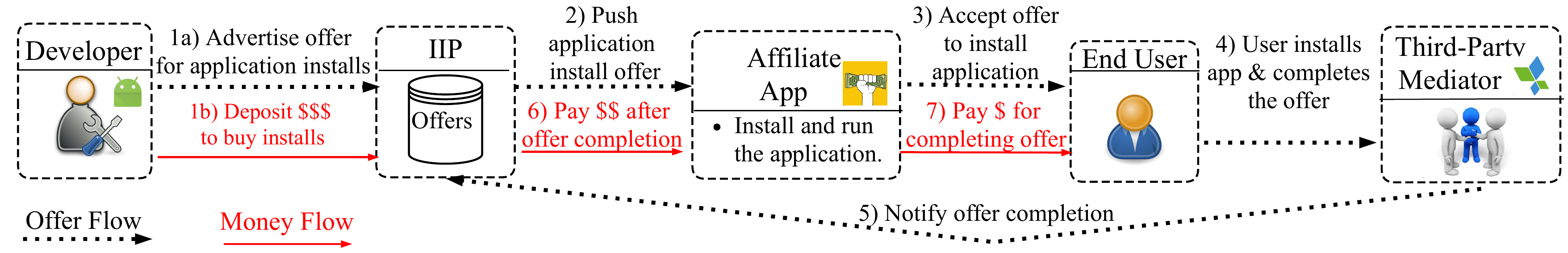}
    \caption{Workflow of an incentivized install campaign. An app developer runs an incentivized install campaign by paying an \incentservice to advertise an offer which is then distributed through an affiliate app where users browse offers and select an offer to work on. Once a user completes an offer (and completion has been certified by a third-party mediator), the payment is disbursed to all the stakeholders.}
    \label{fig: ppinetworkflow}
\end{figure*}

 \begin{figure}[!t]
   \centering
   \includegraphics[width=.99\columnwidth]{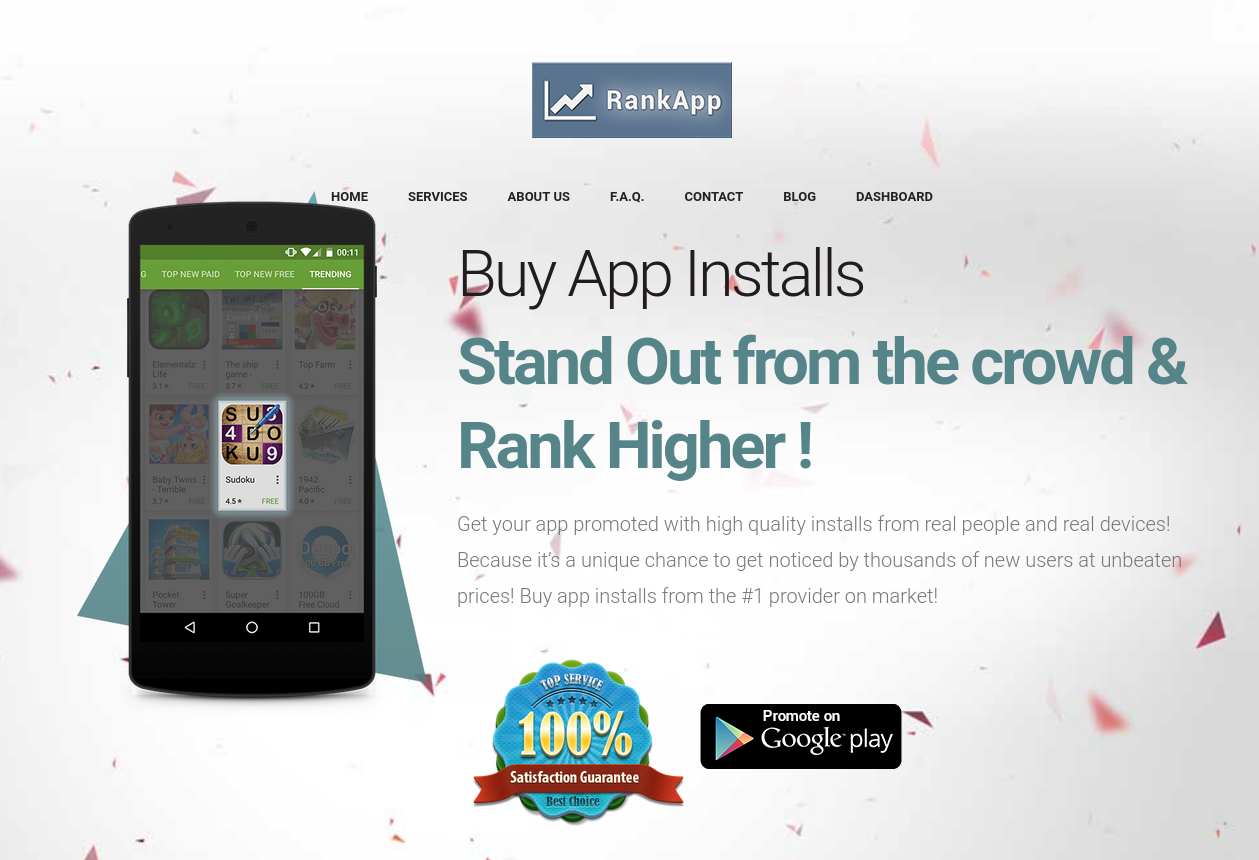}
   \precaption
   \caption{Screenshot of \rankapp showing how they publicly advertise improvement of app's rank on \googlePlay as a service.}
   \label{fig: rankapp}
\postcaption
\postcaption

\end{figure}

In this section, we explain the ecosystem behind incentivized install campaigns and its dynamics. The ecosystem of incentivized mobile app installs involves developers, incentivized install platforms (\incentservices), affiliate apps, users, and third-party mediators.
Figure \ref{fig: ppinetworkflow} describes the flow of the offer and money in an incentivized install campaign, highlighting the role of different stakeholders.

\presub \subsection{Stakeholders} \label{ssec: stakeholders} 

\noindent \textbf{App developer.\xspace}
A mobile app developer (or a third-party like marketing companies hired to improve the usage metrics of an app) runs an incentivized install campaign by paying an \incentservice (\eg \url{fyber.com}, \url{ayetstudios.com}, \url{offertoro.com}) to advertise an offer. The offer contains app details (name and app store URL), a payout, and a description of actions (\eg ``Install and Register,'' ``Install and Reach level 10'') that a user must perform to complete the offer. This offer is then distributed through affiliate apps, where users browse offers and select an offer to work on. 
Once a user completes an offer and the offer completion is certified by a third-party mediator, the payment is disbursed to all the stakeholders. 

\noindent \textbf{Incentivized Install Platforms (IIPs).\xspace}
\incentservices provide an easy and organized way for app developers to manage and run their incentivized advertising campaigns. 
An \incentservice aggregates offers from developers in the form of an \textit{offer wall}. 
These offers are then advertised to a network of affiliate apps that integrate the offer wall of the \incentservice.
\incentservices allow developers to  scale their incentivized advertising campaigns by providing access to  users from different affiliate apps and third-party mediation services.
From the point of view of developers, there is a spectrum of \incentservices depending on their review process, required monetary commitments, and level of compliance with app store policies.


On one end, we find \textbf{vetted \incentservices} (\eg \url{fyber.com}, \url{offertoro.com}, \url{hangmyads.com}) that have a stringent review process to vet developers. 
In most cases, they require developers to provide extensive documentation (\eg valid TAX id, bank account) and make significant upfront monetary commitments (sometimes as high as thousands of dollars).
As a result, we expect vetted \incentservices to be used predominantly by more established developers who can fulfill these requirements.
Moreover, vetted \incentservices tend to stay compliant with app store policies.

On the other end, we find \textbf{unvetted \incentservices} (\eg \url{rankapp.org}, \url{mopeak.com}, \url{cpimobile.com}) that usually do not have a review process to vet developers. 
The barrier to entry is comparatively low since no documentation is required and a developer can pay as little as 20 dollars to start a campaign.
Moreover, unvetted \incentservices appear to put less emphasis on compliance with app store policies. 
For example, \rankapp publicly advertises that their incentivized install campaigns can help an app improve its rank in \googlePlay (see Figure \ref{fig: rankapp}). This may be in violation of \googlePlay's policy against manipulation of app store metrics~\cite{incentInstallPolicyGoogle}.

We are able to easily identify several operational \incentservices by querying Google search with relevant keywords (\eg ``incentivized installs'' and ``mobile offer walls'') and by investigating relevant discussion threads on Reddit and Quora.
By reviewing the websites of seven \incentservices (that we later study in Section \ref{sec:measuringppi}) and also attempting to register as a developer with these \incentservices, we are able to classify them as ``vetted'' and ``unvetted'' as shown in Table~\ref{tbl: IIPs} following the previous definitions.

\begin{table}[!t]
  \centering
  \footnotesize
  \begin{tabular}{llll}
    \hline
    \textbf{\incentservice} & \textbf{Type} & \textbf{Home URL} \\ \hline

    \fyber         & Vetted            & fyber.com \\ 
    \offertoro     & Vetted            & offertoro.com \\
    \adscend  & Vetted            & adscendmedia.com \\
    \hangmyads     & Vetted            & hangmyads.com \\
    \adgem         & Vetted            & adgem.com \\
    \ayet   & Unvetted          & ayetstudios.com \\
    \rankapp       & Unvetted          & rankapp.org \\ \hline
    
  \end{tabular}
  \precaption
  \caption{Characterization of different \incentservices identified in our study by reviewing their websites and attempting to register 
  with them as a developer.}
  \label{tbl: IIPs}
  \postcaption

\end{table}

\noindent \textbf{Affiliate Apps.\xspace}
\incentservices typically rely on multiple affiliate apps for disseminating advertised offers at scale.
An affiliate app signs up with one or more \incentservices to integrate their offer walls through a purpose-specific SDK. 
This enables the users of the affiliate app to access the offers listed in different offer walls.
After a user completes an offer listed in the offer wall, the \incentservice keeps a fraction of the payout and releases the remaining payout to the affiliate app which, in turn, keeps a fraction of the payout and releases the remaining payout to the user.
Affiliate apps offer either non-monetary awards (e.g., advertising these offers as a means to advance in the game) or monetary incentives (e.g., gift cards) to their users for completing these offers.
We primarily focus on affiliate apps that offer monetary incentives in our study.
As we discuss next, disbursement of the payout depends on certification by a third-party mediator.
%

\noindent \textbf{Third-party Mediator.\xspace}
A third-party mediator (or attribution service) \cite{thirdpartymediator} is an entity trusted both by the developer and the \incentservice.\footnote{Some \incentservices also offer attribution services to track offer completion. However, developers may opt for third-party mediators to reduce the risk of fraud.}
The advertised app integrates a purpose-specific SDK of a third-party mediator (\eg \url{appsflyer.com}, \url{kochava.com}, \url{adjust.com}) to track offer completion. Many of these services also offer analytics and anti-fraud products. 
This mediator charges the developer either a fixed amount or a per-user rate. For instance, \url{appsflyer.com} charges 0.03 USD/user.

\presub \subsection{Offer Types} 

We divide offers in two categories based on the set of actions required from a user to complete an offer.

\begin{itemize}[leftmargin=*]
\item \Installonly offers require users to only install and open the advertised app (\eg ``Install and Launch''). This type of incentivized installs can help an app developer to manipulate app store metrics (\eg install counts, visibility).
However, \installonly offers are unlikely to manipulate other user engagement metrics (\eg average user session length) because a user who only installs the app and opens it once (as required by the task) does not contribute to an increase in session length (discussed in Section \ref{sec: activemeasurement}).

%

\item \Activity offers require users to perform additional in-app tasks in the advertised app after installation. These tasks allow developers to manipulate additional user engagement metrics and require a non-trivial amount of time and effort from users. 
A developer can create \activity install offers to inflate the number of registered users (\eg ``Install and Register''), user session lengths (\eg ``Install and Reach Level 10''), and revenue (\eg ``Install and make a \$4.99 in-app purchase'').
These actions result in artificially generated user engagement that could lead to monetization opportunities (discussed in Section \ref{sec:measuringppi}). 
For instance, requiring users to spend a longer amount of time in the app allows monetization through in-app advertising and increases the metrics related to user session length. 
Likewise, inflated user engagement metrics (\eg active users) might help developers raise funding from VCs.
\end{itemize}

\presec \section{Measurements of Incentivized Install Platforms} \postsec
\label{sec: activemeasurement}
In this section, we study \incentservices by purchasing incentivized installs from them.
These measurements not only allow us to get the first-hand experience of \incentservices as a mobile developer but also help us measure the quality of installs (\eg user engagement and retention).
Finally, these measurements also shed light on Google Play's existing enforcement to detect these incentivized installs.
Next, we discuss our purpose-built \honeyapp designed to purchase incentivized installs from \incentservices.


\presub \subsection{Honey Mobile App} \postsub
\label{ssec:honeyapp}
We customize an open-source ``voice memos saving'' Android app and publish it on the \googlePlay to serve as our \honeyapp. 
We add instrumentation to upload metadata to our server. Specifically, our honey app collects information about user in-app activity (\eg clicks on voice memo record button) and device information (\eg list of other installed apps, device build,\footnote{We look for strings (\eg generic, genymotion) to detect emulators.} WiFi SSIDs, the /24 block of the public IPv4 address, and signals to identify whether the device is rooted\footnote{We use the open source library RootBeer \cite{rootBeer}.}).
This information is uploaded to our server whenever the user opens our \honeyapp or clicks the voice memo record button.
This information helps us measure user engagement and characteristics of users who install our \honeyapp. 
We complement our information with the analytics provided by \googlePlay's developer console.

\noindent \textbf{Ethics.\xspace} We received approval from our Institutional Review Board (IRB) before conducting experiments using our honey app.
In our experiment, we try to adhere to the proposed guidelines in the Menlo report \cite{bailey12MenloSP}. 
However, we note that obtaining direct consent from users would not have been feasible since it can influence user actions and findings. Nevertheless, we provide a clear privacy policy to inform users of the nature and purpose of our data collection.
We do not collect device information that might uniquely identify users (\eg we do not collect the IMEI/IMSI of the mobile device, and we drop the last octet 
of the IPv4 address). 
\shehroze{While we do collect the wireless network's SSID, which can be considered personal data, we only store a hashed value. This is enough for our purpose of identifying whether several phones are connected to the same network.}
We acknowledge that while the collection of installed packages in user devices can have privacy risks, it allows us to identify affiliate apps that integrate \incentservices (as we discuss later in Section \ref{sec:measuringppi}) and further understand users' objectives. 
We also take security measures to protect the collected data, and communication with our server happens over encrypted channels.
We only use the collected data for the purpose of our research.
We also received approval from our institution to pay for \incentservices and share billing information (\eg tax ID) when hiring the services of vetted \incentservices.


\presub \subsection{Measuring the Install Campaigns} 
\label{ssec:purchasing}
We arbitrarily pick one vetted (\fyber) and two unvetted (\ayet and \rankapp) \incentservices (from Table \ref{tbl: IIPs}) and  purchase 500 \installonly installs for our \honeyapp.
Our incentivized install campaigns across these three \incentservices are spread over time such that no two campaigns deliver installs at the same time.
We use analytics provided by \googlePlay's developer console \cite{gplayDevStats} to measure the delivery of installs by each \incentservice. Note that we use \googlePlay's developer console to verify that we do not receive any organic installs (\ie installs from app store search or top charts) during our incentivized install campaigns. Thus, we can confidently attribute installs received during each campaign to its respective \incentservice.
%

We combine the information collected by our \honeyapp with analytics from \googlePlay's developer console to analyze user acquisition and user engagement:

\begin{itemize}[leftmargin=*]
    \item \textbf{User acquisition:\xspace} 
    Our \honeyapp received a total of 1,679 installs, including 626 from \fyber, 550 from \ayet, and 503 from \rankapp based on \googlePlay's developer console analytics. The completion time of the campaigns varies between \incentservices. While \fyber and \ayet are able to deliver all the installs within two hours of the campaign launch, \rankapp takes more than 24 hours to deliver all the installs.
    The number of installs for \fyber and \ayet recorded by our server match Google Play's developer console analytics. However, our server is missing telemetry from 45\% (226 out of 503) of the installs purchased from \rankapp. This indicates that a large fraction of \rankapp users likely did not  open the app since our server will not receive telemetry data from an install unless the app is opened.  
    \item \textbf{User engagement:\xspace}
    We analyze user engagement beyond the completion of our advertised offer that only required users to install and open the app to earn the reward.
    Specifically, we measure how many users click on the voice memo record button (\ie the only functionality in the app). 
    Overall, we observe extremely low engagement. In general, less than half of the users clicked on the recording button. When looking at user engagement across \incentservices, we observe lower engagement in \rankapp users: 6\% of \rankapp users clicked on the record button vs. 44\% of \fyber and \ayet users. Nevertheless, user engagement quickly fades over time.  One day after installation, only a handful of users (4 \fyber, 2 \rankapp, 1 \ayet) clicked the record button.

\end{itemize}

Our analysis shows that most of the users who install our \honeyapp do the bare minimum effort to complete the offer. This reaffirms our initial hypothesis that most users are likely not genuinely interested in the apps advertised in the offer walls of \incentservices. Instead, the majority seem to install the \honeyapp to receive the incentive (payout).
 

\noindent \textbf{Incentivized Users.}
Users who install our \honeyapp could be bots who are installing advertised apps from the offer walls in an automated way (similar to click fraud \cite{haddadi10ClickFraud}).
We find that some of our installs show signs of automation.
Our \honeyapp is installed 4 times on an emulator (2 from \rankapp and 2 from \fyber) and 7 of the devices that install our \honeyapp (2 from \fyber, 4 from \ayet, 1 from \rankapp) connect from ASNs of popular cloud services (\eg Digital Ocean) when eyeball ASNs would be expected. 
We also find cases where users seem to deploy device farms \cite{deviceFarmVice} with presumably rooted devices to scale their operations for earning money. 
For example, we record 20 installs from different devices behind the same /24 block. 18 out of these 20 installs are from rooted phones that also share the same WiFi SSID.

We also analyze 17,454 apps that are installed on the devices of potentially real users who installed our \honeyapp to identify their objectives.
The most popular apps among the users of all three \incentservices include several affiliate apps that integrate offer walls from \incentservices.
The most popular affiliate app among the users of \rankapp is \texttt{eu.gcashapp} (37\% users), \ayet is \texttt{com.ayet.cashpirate} (20\% users), and \fyber is \texttt{proxima.makemoney.android} (9\% users).
Our manual analysis shows that the names of many  apps contain keywords such as  ``money'', ``reward'', or ``cash''. 
We find that a large fraction of \rankapp (98\%), \ayet (72\%), and \fyber (42\%) users have installed at least one affiliate app that contains such keywords.
This indicates that most of the users are likely semi-professional crowd workers who seek to earn money through these schemes. 
As we discuss later, we instrument a set of these affiliate apps to identify other mobile apps that run incentivized install campaigns on \incentservices.

Note that we do not purchase installs from more \incentservices due to our limited budget.
Thus, our results may lack completeness. 
Nevertheless, our selection contains at least one \incentservice from the vetted and unvetted categories to reflect the dynamics of both types of \incentservices in our results.

\noindent \textbf{Takeaway.}
Our measurements of installs provided by \shehroze{three} \incentservices suggest that many of the users were likely not genuinely interested in our app and were rather crowd workers installing apps advertised on \incentservices to earn money.
These installs did increase the public install count of our \honeyapp from 0 to 1,000, which indicates that they are effective for manipulating an app's \googlePlay install count metric. This demonstrates a lack of enforcement by
\googlePlay to detect incentivized installs.


\presec \section{Measuring Incentivized Install Campaigns in the Wild} \label{sec:measuringppi} \postsec
In this section, we study incentivized install campaigns of mobile apps in the wild. 

\begin{enumerate}[leftmargin=*]
\item We develop a monitoring infrastructure to find mobile apps running incentivized install campaigns on \incentservices (Section \ref{ssec:data_collection}).

\item We characterize incentivized install campaigns across different \incentservices in terms of: (1) how much users are being paid to complete different types of incentivized offers, and (2) which mobile apps are being advertised (Section \ref{ssec:ecosystem}).
\item We aim to understand how different types of campaigns could benefit developers in terms of improving (1) user engagement and app store visibility, (2) monetization, and (3) likelihood to raise funding (Section \ref{ssec:impact}).
\end{enumerate}

\presub \subsection{\incentservice Monitoring  Infrastructure} \postsub
\label{ssec:data_collection}

To study \incentservices in the wild, we cannot rely on our \honeyapp because it does not allow us to monitor incentivized install offers for other mobile apps.
To this end, we can either directly sign up with \incentservices using a purpose-built honey affiliate app or instrument real-world affiliate apps.
We pick the latter option as it is more scalable: it allows us to monitor many different \incentservices integrated by various affiliate apps.
Next, we explain our approach to identify and instrument real-world affiliate apps.

\begin{figure}[t!]
  \centering
      \includegraphics[width=.98\columnwidth]{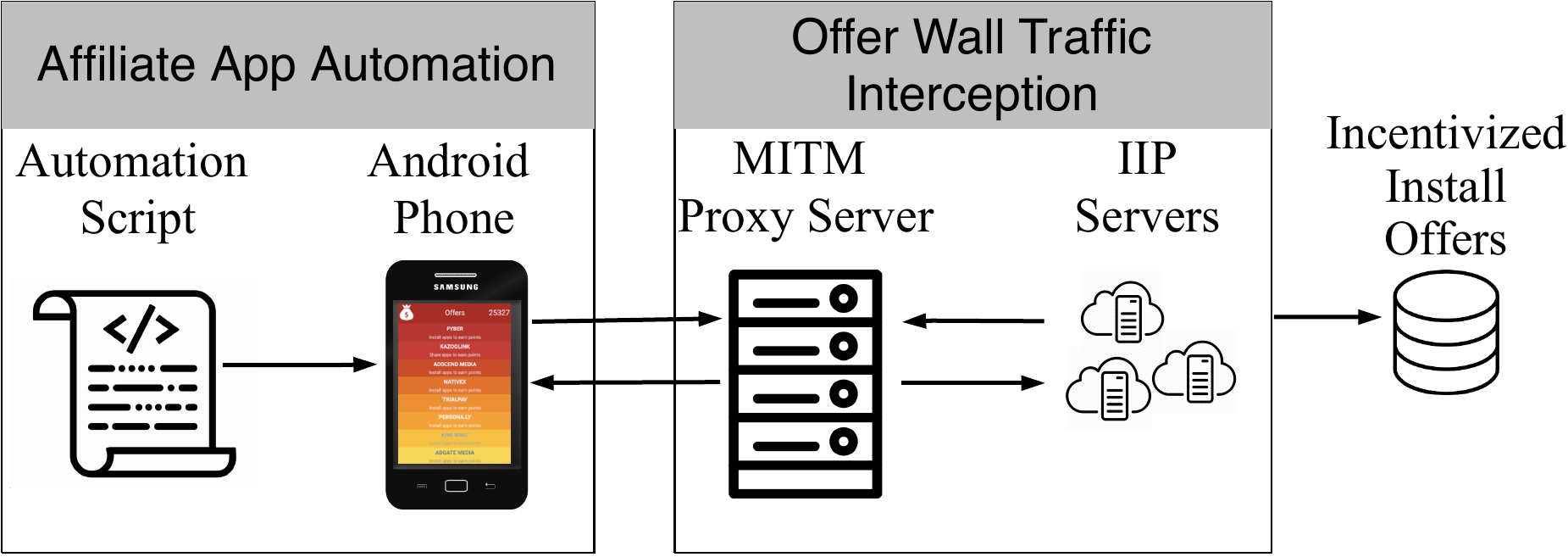}
      \label{fig: milkingProcess}
    \caption{Infrastructure to monitor apps advertised on \incentservices.}
  \label{fig: monitoring}
  \postcaption
\end{figure}

\noindent \textbf{Instrumentation of Affiliate Apps.}
We instrument affiliate apps to extract offer walls of \incentservices that are integrated inside these apps.
Figure \ref{fig: monitoring} shows the two main  components of our instrumentation:

\begin{enumerate}[leftmargin=*]

\item We implement a UI fuzzer based on Appium \cite{appium} to automate UI interactions with an affiliate app on an Android phone. This allows us to milk the affiliate apps without any human in the loop. We note that each affiliate app may have multiple offer walls that are organized in separate tabs. Our UI fuzzer sequentially opens all of the tabs to load the offer walls and then it scrolls through the offer wall to make sure that all the offers are loaded.

\item We intercept the network traffic generated by offer walls on the Android phone by configuring a mitmproxy \cite{mitm} proxy server.\footnote{Note that all offer walls use TLS encryption in their traffic. We decrypt this traffic by installing a self-signed certificate on the Android phone since none of the offer walls uses certificate pinning.}
Specifically, we parse the HTTP responses that are intercepted by the mitmproxy as a result of the stimuli generated by our UI fuzzer when loading an offer wall.
These responses typically include offer details in JSON format containing offer description, payout, and the advertised app's \googlePlay profile.

\end{enumerate}

From the list of affiliate apps identified in Section \ref{sec: activemeasurement}, we select 8 affiliate apps that are available on \googlePlay and are fairly popular.
Using the above methodology, we instrument these 8 affiliate apps as listed in
Table \ref{tbl: affiliateAppsList}.
These affiliate apps have millions of installs, with as many as 10M+ installs for the most popular app.
Note that we do not automate more affiliate apps since it entails significant engineering effort.
We acknowledge that our instrumentation of 8 affiliate apps to monitor \incentservices may lack completeness.
Nevertheless, we are able to identify advertised apps from the offer walls of 7 \incentservices integrated inside these 8 affiliate apps.
Most of the instrumented affiliate apps integrate multiple \incentservices, with the most popular one integrating offer walls from 4 different \incentservices.
We note that all of the 8 affiliate apps integrate at least one offer wall from vetted \incentservices (\fyber, \hangmyads, \adscend, \adgem, 
\offertoro), but most (5 out of 8) of them also integrate at least one offer wall from unvetted \incentservices (\ayet, \rankapp).
\shehroze{
We further increase our coverage of apps advertised on the monitored \incentservices by running our milkers from the following eight countries: USA, UK, Spain, Israel, Canada, Germany, India, and Russia using datacenter VPN proxies offered by \url{luminati.io}.
}





\begin{table}[!t]
  \centering
  \footnotesize
  \begin{tabular}{lrccccccc}
      \textbf{App Package Name} & \textbf{Installs} & \multicolumn{1}{c}{\rotatebox{90}{\textbf{\fyber}} }& \rotatebox{90}{\textbf{\adscend} \smallskip } & \rotatebox{90}{\textbf{\adgem}} & \rotatebox{90}{\textbf{\offertoro}} & \rotatebox{90}{\textbf{\hangmyads}} &  \multicolumn{1}{|c}{ \rotatebox{90}{\textbf{\rankapp}}} &  \multicolumn{1}{c}{\rotatebox{90}{\textbf{\ayet}}} \\
    \hline
     com.mobvantage. & 10M+ &  \cmark & \xmark & \cmark & \xmark & \cmark & \xmark & \cmark \\
     CashForApps & &  & & & & & &   \\
    
     proxima.makemoney & 5M+ &  \cmark & \cmark & \xmark & \xmark & \xmark & \xmark & \xmark \\
         .android & &  & & & & & &  \\
    proxima.moneyapp & 1M+  & \cmark & \xmark & \xmark & \xmark & \xmark & \xmark & \xmark \\
     .android & &  & & & & & &  \\
     com.bigcash.app & 1M+  & \xmark & \cmark & \xmark & \cmark & \xmark & \xmark & \xmark \\
     com.ayet.cashpirate & 1M+ &  \cmark & \xmark & \xmark & \xmark & \xmark & \xmark & \cmark \\
      eu.makemoney& 1M+ &  \xmark & \cmark & \xmark & \xmark & \xmark & \cmark & \xmark \\
      com.growrich. & 1M+ & \xmark & \cmark & \xmark & \xmark & \xmark & \cmark & \xmark \\
      makemoney & &  & & & & & & \\
    make.money.easy & 100K+ & \cmark & \cmark & \xmark & \xmark & \xmark & \xmark & \cmark\\
     \hline
  \end{tabular}
  \precaption
  \caption{List of monitored affiliate apps and the offer walls of \incentservices integrated
   inside them. We instrument these eight affiliate apps to find 
	 \googlePlay apps advertised on vetted and unvetted \incentservices. \rankapp and \ayet are unvetted \incentservices while the remaining services are vetted \incentservices.}
\label{tbl: affiliateAppsList}
\postcaption

\end{table}

\noindent \textbf{Dataset.}
We use our instrumentation of affiliate apps to monitor \incentservices over a period of three months from March-June 2019.
We identify a total of 2,126 offers from 922 unique advertised apps that are advertised on 7 \incentservices, both vetted and unvetted services. 
%
We note that offer payouts use different point systems across different affiliate apps.
We normalize offer payouts of different affiliate apps by converting their points to equivalent dollar amounts.\footnote{By analyzing affiliate apps, we convert these reward points 
to an equivalent offer payout in USD that can be redeemed through gift cards 
(\eg PayPal, Amazon)
inside the affiliate app.}

Overall, we identify a total of 1,128 unique offer descriptions in these 2,126 offers.
We manually label offer descriptions into two offer types (\installonly and \activity) according to the classification described earlier in Section \ref{sec: incentecosystem}.
Our manual analysis of these offer descriptions further reveals that some \activity offers are targeting specific user engagement metrics.
Hence, we further divide \activity offers into three subcategories: 
(1) \textit{\Register} if the offer requires users to register an account, 
(2) \textit{\Apppurchase} if the offer requires users to make in-app purchase, and
(3) \textit{\Appusage} if the offer requires users to perform any other action.

\begin{table}[!t]
    \centering
    \footnotesize
      \begin{tabular}{lcc}
        \hline
        \textbf{Offer Type} & \textbf{\% of offers ($N=2,126$)} & \textbf{Average payout} \\ 
        \hline
        \Installonly & 47\% &  \$0.06 \\ 
        \Activity & 53\% & \$0.52  \\ 
        \Activity (\Appusage) & 37\%  &  \$0.50  \\ 
        \Activity (\Register) & 11\% &  \$0.34  \\ 
        \Activity (\Apppurchase) & 5\% &  \$2.98 \\ 
        
        \hline
      \end{tabular}
      \precaption
      \caption{Prevalence of different types of incentivized install offers and their average payouts.}
    \label{tbl: installTypeStats}
    \postcaption
    
\end{table}

\begin{table*}[!t]
  \centering
  \footnotesize
    \begin{tabular}{ccrrrrrrrrr}
      \hline
      \textbf{\incentservice} &\textbf{\incentservice} & \textbf{Median} & \multicolumn{2}{|c|}{\textbf{Offer Type}} & \textbf{Number} & \textbf{Number} & \textbf{Number} & \textbf{Number} & \textbf{Median} &  \textbf{Median Age} \\ 
      
      \textbf{Name} &\textbf{Type} &\textbf{Offer} &  \multicolumn{2}{|c|}{\textbf{}} & \textbf{of} &  \textbf{of} & \textbf{of} & \textbf{of} & \textbf{Install}  & \textbf{of Apps} \\ 
      
       &  & \textbf{Payout} & \multicolumn{1}{|c}{\textbf{\Installonly}} &  \multicolumn{1}{c|}{\textbf{\Activity}} & \textbf{Apps} & \textbf{Developers} & \textbf{Countries} & \textbf{Genres} & \textbf{Counts} &  \textbf{(Days)} \\ 
      
      \hline

      \textbf{\rankapp} & Unvetted & \$0.02 & 100.0\% & 0\% & 152 & 114 & 39 & 20 & 100  &  33 \\ 
      \textbf{\ayet} & Unvetted & \$0.05 & 71\% & 29\% &      392 & 351 & 44 & 51 & 1K   &  70 \\ 
      \textbf{\fyber} & Vetted & \$0.19 & 24\% & 76\% &     378 & 319 & 40 & 36 & 1M   &  777 \\ 
      \textbf{\adscend} & Vetted & \$0.12 & 9\% & 91\% &   104 & 79  & 27 & 21 & 500K &  722 \\ 
      \textbf{\adgem} & Vetted & \$1.71 & 16\% & 84\% &     28  & 27  & 15 & 8 & 500K &  854  \\ 
      \textbf{\hangmyads} & Vetted & \$0.40 & 23\% & 77\% & 27  & 27  & 17 & 9 & 1M   &  699 \\ 
      \textbf{\offertoro} & Vetted & \$0.09 & 52\% & 48\% & 140 & 131 & 34 & 19 & 500K &  557 \\ 
    
        \hline
     
    \end{tabular}
    \precaption
    \caption{Summary of \googlePlay's metadata and offers' metadata of apps identified on vetted and unvetted \incentservices. Mobile apps from a wide variety of app genres advertised on several \incentservices are published on \googlePlay by hundreds of developers based in different countries. We also note that less-established (newer, less popular) apps are more likely to run \installonly incentivized install campaigns and use unvetted \incentservices rather than vetted \incentservices presumably because of their higher barrier of entry.}
    
  \label{tbl: offerwallCharacterisitc}
  \postcaption
 \end{table*}

\presub \subsection{Characterizing Incentivized Install Offers} \postsub
\label{ssec:ecosystem}
We analyze offers and advertised mobile apps across different vetted and unvetted \incentservices.

\noindent \textbf{Offer Types and Payouts.}
Table \ref{tbl: installTypeStats} reports the presence of different types of incentivized install offers and their average payouts. 
We observe that \activity offers are slightly more popular than \installonly offers.
53\% are \activity offers while 47\% are \installonly offers.
We observe that the average payout of \activity offers is 9$\times$ higher than that of \installonly offers.
We further observe differences among the payouts of the subcategories of \activity offers.
The average payout of \apppurchase offers is 9$\times$ and 6$\times$ higher than that of \register and \appusage offers, respectively.
This is not surprising because users have to put in more effort and even spend money (to make in-app purchases) for completing \activity offers. 
We conclude that manipulating advanced engagement metrics through \activity offers is generally much more expensive than manipulating simple install counts through \installonly offers. 
%

Table \ref{tbl: offerwallCharacterisitc} shows that there are differences in offer payouts across vetted and unvetted \incentservices.
The lower payouts in unvetted \incentservices are because developers mostly advertise \installonly offers on unvetted \incentservices.
Specifically, one unvetted \incentservice (\rankapp) has no \activity offers and an average payout of \$0.02, and the other unvetted \incentservice (\ayet) has only 29\% \activity offers and an average payout of \$0.05.
In contrast, vetted \incentservices advertise much more \activity offers with higher payouts.
These findings show that incentivized install campaigns on vetted \incentservices are more interested in manipulating advanced engagement metrics.

\noindent \textbf{Advertised Apps and Developers.}
Next, we compare advertised mobile apps and their developers across vetted and unvetted \incentservices.
Table \ref{tbl: offerwallCharacterisitc} lists some basic statistics of advertised mobile apps and their developers by crawling their \googlePlay's profile
to collect metadata such as their install counts (Google reports installs in bins of a lower-bound ``minimum'' number of installs), release date, genre, and developer details (\eg  company name, websites, mailing address, developer ID).
We note that mobile apps from a wide variety of app genres advertised in several \incentservices are published on \googlePlay by hundreds of developers based in different countries.\footnote{Note that we uniquely identify a developer using the developer's ID and identify the developer's country by parsing their mailing addresses from their app's Google Play Store profile. Unique apps are identified by their package name.}
As an example of unvetted \incentservices, 392 apps of 51 genres advertised on \ayet are published on \googlePlay by 351 developers based in 44 countries.
As an example of vetted \incentservices, 378 apps of 36 genres advertised on \fyber are published on \googlePlay by 319 developers based in 40 countries.
It is noteworthy that some of the popular apps are from mainstream developers.
For example, we identify ``Apple Music'' from Apple and ``LinkedIn: Job Search \& Business News'' from LinkedIn on vetted \incentservices. 
Similarly, we identify ``TikTok - Make Your Day'' from TikTok and ``Fiverr - Freelance Services'' from Fiverr on unvetted \incentservices.
It is unclear whether these developers are purchasing incentivized installs themselves or if they are potentially deceived by third parties (e.g., marketing companies) or \incentservices, which run incentivized installs to fulfill regular mobile app install advertising campaigns \cite{appsFlyerAdSpent} (as we discuss later in Section \ref{sec: discussion}).
Table \ref{tbl: offerwallCharacterisitc} shows that there are significant differences in the age of the apps\footnote{We measure the age of an app by computing the difference between the start of the advertised app's campaign date and the date in which the app was released on the \googlePlay.} and the popularity of apps advertised on vetted and unvetted \incentservices.
For example, the median install count of apps advertised on both unvetted \incentservices is 100 and 1,000, respectively, while it ranges 
between 500,000 and 1,000,000 for vetted \incentservices.
As another example, the median age of apps advertised on both unvetted \incentservices is 33 and 70 days, respectively, while it ranges 
between 557 and 854 days for vetted \incentservices.
Overall, we conclude that less-established (e.g., newer or less popular) apps are more likely to use unvetted \incentservices rather than vetted \incentservices presumably because of their higher barrier of entry.
Next, we show how well-established apps differently benefit from running \activity campaigns on vetted \incentservices as compared to less-established apps that generally run \installonly campaigns on unvetted \incentservices.


\presub \subsection{Impact of Incentivized Install Campaigns} 
\begin{figure}
    \centering
    \includegraphics{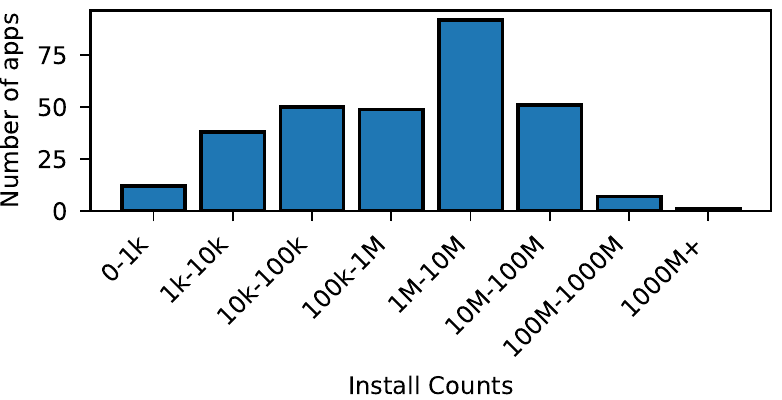}
    \caption{Install counts of the baseline apps.}
    \label{fig:baseline_installs}
    \postcaption
\end{figure}

\postsub
\label{ssec:impact}
In this section, we study how developers benefit from different types of incentivized campaigns on vetted and unvetted \incentservices. First, we measure the impact of incentivized installs on app store metrics.  Second, we analyze how some developers monetize different types of \activity offers in their incentivized campaigns.
Finally, we study whether incentivized installs might help developers raise funding. 
Note that some confounding factors (\eg non-incentivized installs) may have an effect on the advertised apps to improve app store metrics or raise funding.
To mitigate the impact of such factors, we compare the set of apps running incentivized install campaigns with a set of 300 baseline apps published on Google Play. 
Our baseline apps are sampled from the set of apps commonly used by Lumen users~\cite{razaghpanah2015haystack}, a crowd-sourced privacy tool.
Figure \ref{fig:baseline_installs} shows that our baseline contains a comprehensive set of apps of all types of popularity ranging from less than 1K installs to over 1000M installs.
Note that our sample of baseline apps has no overlap with the apps advertised on \incentservices that are identified in our study.
We also use the chi-squared test of independence \cite{chiSquareTest} to determine whether there is a statistically significant difference in improved app store metrics and fund raising between apps observed in incentivized install campaigns, and those in the baseline.
We acknowledge that any correlation we might find in this section does not necessarily imply causation because there could be other confounding factors beyond our purview (e.g., non-incentivized install campaigns).

\presubsub \subsubsection{Impact on App Store Metrics} \postsubsub
\label{sssec:ASO}
We analyze whether incentivized install campaigns from vetted and unvetted \incentservices may be able to manipulate app store metrics.
To this end, we crawl \googlePlay profiles of  apps to collect their install counts.
We also crawl \googlePlay ``top charts'' that list trending apps (\eg top gaming apps, top grossing apps). 
We periodically collect this data every other day from March 2019 to June 2019.
This allows us to measure how IIP campaigns on vetted and unvetted \incentservices help apps increase their install counts and visibility on \googlePlay.

\noindent\textbf{Increase in Install Counts.}
For this analysis, we compare the increase in install counts of  apps advertised on vetted and unvetted \incentservices with our set of baseline apps.
Specifically, we check whether or not an app's install count increases by the end of the incentivized install campaign as compared to the start of the campaign. 
For a fair comparison, we monitor install counts for baseline apps over the duration of 25 days -- which is the average incentivized install campaign duration.
Table~\ref{tbl: installCountsIncrease} compares the number of apps from vetted \incentservices, unvetted \incentservices, and baseline that increased their install counts.
To determine whether incentivized install campaigns significantly impact the increase in install counts, we use the chi-squared test of independence~\cite{chiSquareTest,chiSqExample}. 
We conduct two separate tests, ``vetted vs.\ baseline'' and ``unvetted vs.\ baseline''  to compare vetted and unvetted \incentservices with the baseline.
Our null and alternate hypotheses are:

\begin{itemize}[leftmargin=*]
\item $H_o$: The proportion of apps whose install count increases is independent of incentivized install campaigns.

\item $H_a$: The proportion of apps whose install count increases is associated with incentivized install campaigns.
\end{itemize}

For vetted vs.\ baseline, $\chi^2=26.0$ and $p=3.378e^{-7}$.
For unvetted vs baseline, $\chi^2=39.9$ and $p=0.000$.
Thus, we are able to reject the null hypothesis for both tests at the significance level of $0.05$.
We conclude that there is a correlation between apps advertised on vetted and unvetted \incentservices and an increase in the install counts of advertised apps.
Comparing vetted and unvetted \incentservices, we note that the likelihood of an increase in install counts (with respect to the baseline) for vetted \incentservices is 6$\times$ while it is 8$\times$ for unvetted \incentservices.
%
These findings seem counter-intuitive since apps using unvetted \incentservices, despite cheaper offer payouts as shown in Table \ref{tbl: offerwallCharacterisitc}, are benefiting more than vetted \incentservices.
Nevertheless, as we discuss next, apps that use vetted services benefit in different ways, other than simply increasing their install counts.

\begin{table}[!t]
  \centering
  \footnotesize
    \begin{tabular}{crr}
      \hline
      \textbf{App Set}&    \textbf{No Increase} & \textbf{Increase}\\
      \hline
       Baseline ($N = 300 $)  & 294 (98\%) & 6  (2\%)  \\
      Vetted ($N = 492 $)  & 431 (88\%)  &   61 (12\%)  \\
      Unvetted ($N = 538 $)   & 450 (84\%)  & 88 (16\%)  \\
      \hline
      
    \end{tabular}
    \precaption
    \caption{Comparing apps that increased install counts from vetted and unvetted \incentservices with baseline apps. Through the chi-squared test of independence, we find that there is indeed a correlation between apps advertised on both vetted and unvetted \incentservices and an increase in the install counts of advertised apps.}
    
\label{tbl: installCountsIncrease}
  \postcaption
\end{table}

\begin{table}[!t]
  \centering
  \footnotesize
    \begin{tabular}{crr}
      \hline
      \textbf{App Set} &  \textbf{Not Present} & \textbf{Present}\\
      \hline
       Baseline ($N = 261 $) & 253 (96.9\%) & 8  (3.1\%)  \\
      Vetted ($N = 320 $)    & 296 (92.5\%) & 24 (7.5\%) \\
      Unvetted ($N = 484 $)  & 472 (97.5\%) & 12 (2.5\%)  \\
      \hline

    \end{tabular}
    \precaption
    \caption{Comparing the appearance of advertised apps from vetted and unvetted \incentservices with baseline apps in top charts. Through the chi-squared test of independence, we find that there is a correlation between apps advertised on vetted \incentservices and the appearance of an app in the top chart.} 
    
  \label{tbl: visibility}
  \postcaption
\end{table}

\begin{figure}[t!]
  \centering
      \subfigure[com.mmm.trebelmusic]{ 
      \includegraphics[width=.66\columnwidth]{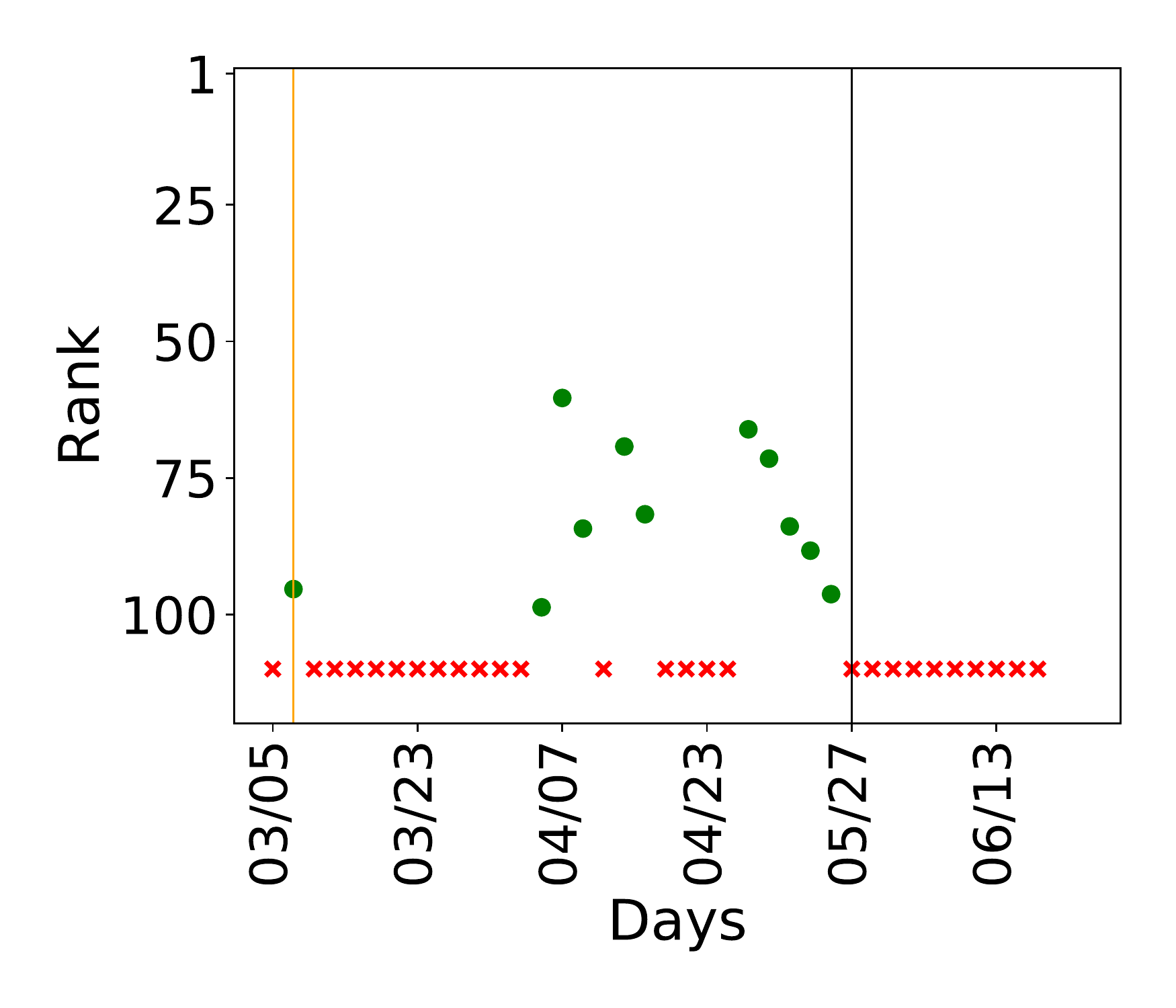}
      \label{fig: TopAso1}
      }
      \subfigure[com.camelgames.wof]{ 
      \includegraphics[width=.66\columnwidth]{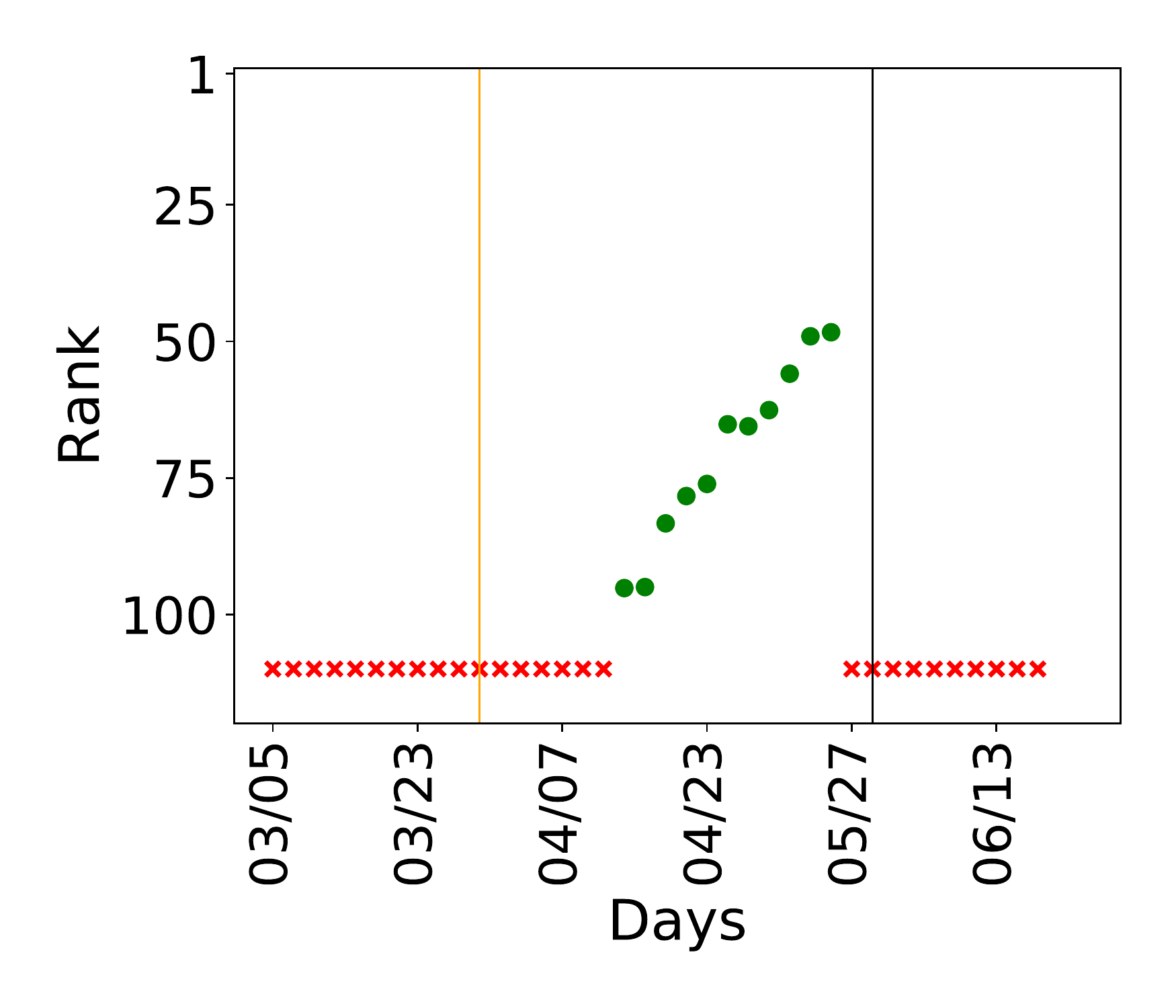}
      \label{fig: TopAso2}
      }
    
  \precaption
  \caption{Case studies of the appearance of advertised apps in top charts during their campaign. The yellow line indicates the start of the campaign and the black line indicates the end of the campaign. Each red cross indicates the days when an app is not observed in a particular chart and each green dot indicates the days when an app has appeared. The y axis shows the percentile rank of the app. 
}
  \label{fig: TopAsoManipulationPlots}
  \postcaption
\end{figure}

\noindent \textbf{Appearance in Top Charts.}
For this analysis, we compare top chart appearances
-- \ie their app store visibility --
of apps advertised on vetted and unvetted \incentservices with our set of baseline apps.
We check whether or not an app appears in the top charts by the end of their install campaign (\ie the last day we observe them in our crawl). 
To minimize bias, we exclude advertised apps that already appeared in top charts before the start of their campaign (\ie the first day we observe them in our crawl) and baseline apps that appeared in top charts at the start of our crawls.
After this filtering, we are left with 320 apps from vetted \incentservices, 484 apps from unvetted \incentservices, and 261 baseline apps.
For a fair comparison, we again monitor visibility for the baseline apps over the duration of 25 days, which is the average incentivized install campaign duration.
Table~\ref{tbl: visibility} reports the number of these apps from vetted \incentservices, unvetted \incentservices, and baseline apps that appeared in top charts.
To determine whether incentivized install campaigns significantly impact app store visibility, we use the chi-squared test of independence. 
%
%
Our hypotheses are:

\begin{itemize}[leftmargin=*]
\item $H_o$: The proportion of apps that appear in top charts is independent of incentivized install campaigns.

\item $H_a$: The proportion of apps that appear in top charts is associated with incentivized install campaigns.
\end{itemize}

For vetted vs. baseline, $\chi^2=5.43$ and $p=0.02$.
For unvetted vs baseline, $\chi^2=0.22$ and $p=0.64$.
Thus, we are able to only reject the null hypothesis for the vetted vs. baseline test at the significance level of $0.05$.
We conclude that there is a correlation between apps advertised on vetted \incentservices and the appearance of an app in top charts.
Mobile apps advertised on vetted \incentservices appear in top charts up to 2$\times$ more frequently in comparison to baseline apps as shown in Table \ref{tbl: visibility}.
However, we are unable to conclude whether advertising apps on unvetted \incentservices affects the chances of appearance in top charts.
Vetted \incentservices not only help apps increase their install counts but can also help them appear in top charts. 
Note that even though developers generally pay more for \activity offers on vetted \incentservices, it appears worthwhile because \googlePlay places apps in top charts based on user engagement metrics \cite{engagementVisibilityGplay}, which cannot be inflated with \installonly offers on unvetted \incentservices.

To further illustrate this point, we analyze apps that use different types of \activity offers as listed in Table \ref{tbl: installTypeStats}.
Figure \ref{fig: TopAsoManipulationPlots} shows the temporal evolution of the rank of two example apps in the top charts. 
In Figure \ref{fig: TopAso1}, the app ``TREBEL - Free Music Downloads \& Offline Play'' starts appearing in the top-games chart after the start of its incentivized install campaign using \register and \appusage offers. 
Specifically, the description of these offers is ``Install and register'' and ``Install, register, and download a song.''
This suggests that manipulation of more targeted user engagement metrics through \register and \appusage offers can help apps appear in top charts.
In Figure \ref{fig: TopAso2}, the app ``World on Fire'' appears in the top-grossing chart a few days after the start of its incentivized install campaign using \apppurchase offers.
Specifically, the description of one of these \apppurchase offers is ``Install \& Make any purchase.''
This suggests that these \apppurchase offers manipulated revenue through fake purchases, which in turn helped ``World on Fire'' appear in the ``Top Grossing Apps'' chart.
We conclude that inflating user engagement metrics (\eg time spent in the app) and revenue through \activity offers may help apps appear in relevant top charts.
As we discuss next, these \activity offers also seem to provide monetization opportunities to developers.

\noindent \textbf{Summary.}
Our results show differences in the effectiveness of running incentivized install campaigns on vetted and unvetted \incentservices.
While unvetted \incentservices seem to be more effective in targeting a primitive engagement metric (\ie install counts), vetted \incentservices can be more effective in manipulating advanced user engagement metrics and app store visibility. 
Our results also show that \googlePlay's detection and filtering systems \cite{spamInstallsGoogleBlog,incentInstallsrankingsGoogleBlog} likely fail to effectively enforce their policy against app store manipulation by incentivized installs \cite{incentInstallPolicyGoogle}.

\presubsub \subsubsection{Monetization} \postsubsub
\label{sssec:revenueManipulation}
We discuss how \activity offers provide different opportunities for monetization through advertisement and arbitrage.

\noindent\textbf{Advertising.}
In-app advertising is a popular way to monetize mobile  apps~\cite{razaghpanah2018apps}.
Intuitively, there is a direct relationship between user engagement and revenue earned from advertisements. 
Higher user engagement provides developers more opportunities to show ads, something known as ``active eyeballs'' in Ad Tech jargon~\cite{eyeballAds}.
Many \activity offers require users to engage with the app for a long time period by asking them to complete a series of difficult tasks such as ``Install \& Reach level 10''. 
This user engagement can be directly monetized by displaying in-app ads to the user.
As a proxy for measuring advertising revenues, we analyze whether apps using \activity offers have more capability to monetize through advertising libraries. 
To this end, we analyze the number of advertisement libraries embedded in apps.
We download APKs of baseline and advertised apps to perform static analysis\footnote{We note that static analysis may miss some advertising libraries due to code obfuscation and dynamic code loading. Furthermore, we acknowledge that our analysis offers an upper bound on in-app advertising, as the mere presence of an advertising library does not guarantee that its code will be executed at run-time.} using LibRadar~\cite{ma2016libradar}.

\begin{figure}[!t]
  \centering
  \subfigure[Offer activity type]{
  \includegraphics[width=.95 \columnwidth]{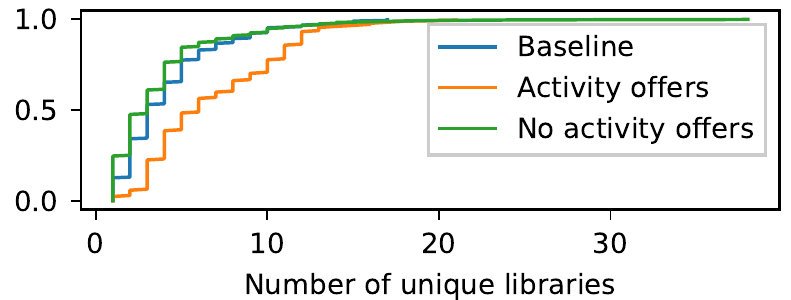}
  \label{fig: adLibs_offers}
  }
  \subfigure[IIP type]{
  \includegraphics[width=.95 \columnwidth]{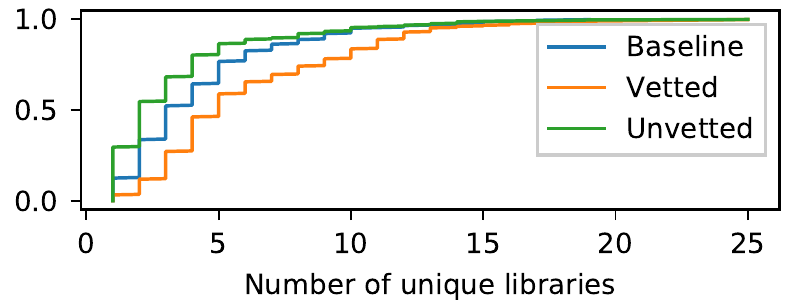}
  \label{fig: adLibs_iip}
  }
  \precaption
  \caption{Distribution of unique ad libraries across apps.}
  \postcaption
\end{figure}

Our results show that apps using \activity offers have more ad libraries than apps using \installonly offers.
Figure \ref{fig: adLibs_offers} shows that 60\% and 25\% of apps with \activity and \installonly offers, respectively, have 5 or more ad libraries.
We find well-known advertising vendors such as Google AdMob, AppLovin, and ChartBoost.
We also find advertisers that serve the role of \incentservice (\eg Fyber).
A similar pattern emerges when comparing vetted, unvetted, and baseline apps. 
Figure \ref{fig: adLibs_iip} shows that 55\%, 20\%, and 35\% of vetted, unvetted, and baseline apps, respectively, have 5 or more ad libraries.
This is expected because \activity offers are more common on vetted \incentservices in comparison to unvetted (as reported in Table \ref{tbl: installTypeStats}).

Our results indicate two different strategies of incentivized install campaigns. 
On one hand, we find that mobile apps advertised on vetted \incentservices have more advertisement libraries, giving these apps  more opportunities to generate revenue since their incentivized install campaigns are more likely to advertise \activity offers.
On the other hand, incentivized install campaigns from unvetted \incentservices advertise \installonly offers that do not require additional engagement from users and include fewer advertising libraries.

\noindent\textbf{Arbitrage.}
We observe that developers can craft \activity offers in a way that allows them to monetize through arbitrage.
In arbitrage, a developer advertises an \activity offer that requires users to complete additional tasks within the app such as filling surveys, watching videos, or even installing other apps (similar to affiliate apps).
In return, the developer earns money through commissions after these offers within the app are completed.
One example of arbitrage is the following offer that advertises an app ``Cash Time - Earn Money - Android''.
This \activity offer pays \$0.67 to users for reaching 850 points in the app by completing additional tasks (e.g., watching videos, completing surveys, shopping for deals) in the app.
Unfortunately, we cannot estimate the revenue earned through arbitrage to understand profitability. However, even if arbitrage is not profitable, these offers would likely enable an app to increase revenue-based metrics, such as revenue growth.
We manually analyze the descriptions of all \activity offers and find that 3.9\% of apps (36 out of 922) use arbitrage-based \activity offers.
We observe a higher percentage of such offers in vetted \incentservices as compared to unvetted \incentservices.
Specifically, 7\% of apps (35 out of 492) from vetted \incentservices while only 2\% of apps (10 out of 538) from unvetted \incentservices use arbitrage-based \activity offers. 
This is again expected because \activity offers are more common on vetted \incentservices in comparison to unvetted (as reported in Table \ref{tbl: installTypeStats}).

\noindent \textbf{Summary.}
Overall, our results show that incentivized install campaigns that advertise \activity offers have opportunities to monetize through advertising and arbitrage.
It is unclear whether these monetization strategies are sufficient to directly recuperate the cost of their incentivized install campaigns. 
Next, we explore other indirect avenues of monetization such as raising funding from venture capitalists.

\presubsub \subsubsection{Investor Funding}\label{sssec:vcfunding} \postsubsub
There is an inherent tension between growth and profitability \cite{growthVsProfit} of mobile app startups.
Many app developers are more interested in strong user growth as well as bolstering user engagement metrics because, in the long term, they can help developers raise funding from investors such as venture capitalists (VCs) \cite{investorUser,investorInstallsStartups}.
In this section, we study whether any of the 922 apps in our dataset are able to raise funding after launching incentivized install campaigns.
To this end, we use the Crunchbase \cite{crunchbase} database that provides us with access to the list of companies that have raised funding.
We download a snapshot of the Crunchbase database in October 2019 (a few months after our initial data collection).
By searching for developer information from \googlePlay, we match 23\% of 922 apps to their developers in the Crunchbase database. 
Crunchbase provides funding information such as the name of the funding organization, type of organization (\eg angel investor, VC investor), amount of funding, and type of funding (\eg seed round, series A, series B).
We are able to match 39\% of apps (192 out of 492) from vetted \incentservices, 15\% of apps (79 out of 538) from unvetted \incentservices, and 27\% of baseline apps (82 out of 300) in the Crunchbase database.
We surmise that mobile apps advertised on vetted \incentservices have a higher percentage of matched developers over unvetted \incentservices since vetted \incentservices are used by more established developers (as discussed in Section~\ref{ssec:data_collection}).
In fact, our manual analysis shows that mobile apps of the unmatched developers from unvetted \incentservices often do not provide useful information in their \googlePlay profile (\eg link to their website).
Note that the lack of completeness of Crunchbase database may also result in the exclusion of some unmatched developers.
To mitigate the impact of this limitation, we only compare those advertised apps and baseline apps that are matched in the Crunchbase database to perform further analysis.


We analyze whether advertising apps on vetted and unvetted \incentservices has any effect on the funding raised by them.
To this end, we compare the funding raised by developers whose mobile apps are advertised on vetted and unvetted \incentservices with baseline apps.
Specifically, we check whether or not a mobile app raises funding after running the incentivized install campaign(s).
For comparison, we extract the same information for our baseline apps. 
Table~\ref{tbl: fundingCrunchBaseResults} reports the number of these mobile apps from vetted \incentservices, unvetted \incentservices, and from the baseline whose developers are able to raise funding.
To determine whether incentivized install campaigns significantly impact the funding raised by apps, we again use the chi-squared test of independence.
We conduct two separate tests, ``vetted vs.\ baseline,'' and ``unvetted vs.\ baseline,''  to compare vetted and unvetted \incentservices.
Our hypotheses are:

\begin{itemize}[leftmargin=*]
\item $H_o$: The proportion of apps that raise funding is independent of incentivized install campaign.
\item  $H_a$: The proportion of apps that raise funding is associated with incentivized install campaign.
\end{itemize}

For vetted vs.\ baseline, $\chi^2=4.7$ and $p=0.03$.
For unvetted vs.\ baseline, $\chi^2=2.8$ and $p=0.10$.
Thus, we are able to only reject the null hypothesis for the vetted vs.\ baseline at the significance level of $0.05$.
We conclude that there is a correlation between advertising apps on vetted \incentservices and developers of advertised apps raising funding.
However, we are unable to conclude whether advertising apps on unvetted \incentservices affects the chances of raising funding.

We further compare the offer types and payout amounts of incentivized install campaigns for the apps that raise funding versus those that do not. 
%
Table \ref{tbl: fundedvettedapps} presents a breakdown of offer types and payouts of 30 apps from vetted \incentservices that raised funding after their campaigns.
We observe that these apps are roughly equally likely to advertise \activity and \installonly offers.
However, the incentivized install campaigns of the apps that raised funding have much higher payouts.
For example, the average payout of \installonly offers advertised by apps that raised funding is twice the average payout of all offers reported in Table \ref{tbl: installTypeStats}.
We surmise that the developers interested in raising funding need to aggressively acquire new users, and thus are willing to pay more.

We also look at two case studies of apps that raise VC funding after incentivized install campaigns. 
First, the \textit{Password Manager} app by Dashlane raised multiple rounds of funding after launching incentivized install campaigns.
The app was advertised on \fyber{} from March 12-27 and then May 7-14 using \activity offers, which require users to install the app, create an account, and save at least two passwords in return for a reward around \$0.26-\$0.66.
Coincidentally, Dashlane raised funding twice, each time about two weeks after each campaign: \$30 million on April 12 \cite{dashlane-1}, and \$110 million on May 30 \cite{dashlane-2}.
Second, \textit{Droom: Used \& New Car, Bike, Insurance, Loan \& RTO}, an e-commerce app that sells used and new vehicles, raised series F funding after appearing on \incentservices. 
The app was advertised on \fyber and \adscend from March 14-June 8 using \installonly and \activity offers, which require users to install the app and register a new account in return for a reward around \$0.02-\$0.06.
Droom also raised funding about two weeks after its incentivized install campaign: Droom raised \$10 million on 20th June \cite{droom}.
These two case studies indicate that incentivized installs could have misled investors who funded these apps (or the companies behind them). 
In the same vein, incentivized install campaigns of the remaining apps from vetted \incentservices could also have misled investors since they also raised funding after the start of their incentivized install campaigns.


\begin{table}[!t]
  \centering
  \footnotesize
    \begin{tabular}{crr}
      \hline
      \textbf{App Set} &  \textbf{Funding Raised} & \textbf{No Funding Raised}\\
      \hline
       Baseline ($N = 82$) &  5 (6.1\%) & 77 (93.9\%)    \\
      Vetted ($N = 192$)    & 30 (15.6\%)  & 162 (84.4\%) \\
      Unvetted ($N = 79$)  & 11 (13.9\%)  & 68 (86.1\%)  \\
      \hline
      
    \end{tabular}
    \precaption
    \caption{Developers of mobile apps raising funding after campaigns using vetted and unvetted \incentservices compared with baseline apps.
    Through the chi-squared test of independence, we find that there is a correlation between apps advertised on vetted \incentservices and successfully raising funding after a campaign.}
  \label{tbl: fundingCrunchBaseResults}
\end{table}

\begin{table}[!t]
  \centering
  \footnotesize
    \begin{tabular}{crr}
      \hline
      \textbf{Offer Type} &  \textbf{Percentage of Apps $(N=30)$} & \textbf{Average Payout}\\
      \hline
       \Installonly & 67\% & \$0.12 \\
       \Activity & 63\% & \$0.92 \\
      \hline
      
    \end{tabular}
    \precaption
    \caption{Breakdown of offer types and payouts of apps advertised on vetted \incentservices that raised funding after their campaign.}
  \label{tbl: fundedvettedapps}
\end{table}

Next, we study developers matched with the Crunchbase database that are public companies \cite{defPublicComp}.
Public companies generally report their user growth and engagement metrics on a quarterly basis. 
These metrics often play an important role in stock price targets from analysts and eventual stock price movements~\cite{stockPriceEarning}.
Thus, manipulation of these metrics through \incentservice campaigns can potentially boost the stock price of a publicly traded company.
From the matching of developers from the Crunchbase database, we find that developers of 28 advertised mobile apps from vetted and unvetted \incentservices are publicly traded companies. 
For example, Redfin, a publicly traded real estate brokerage company, reported a 27\% increase in their mobile traffic during the second quarter of 2019 in comparison to the second quarter of 2018 \cite{redfinQ2}. 
Note that the incentivized install campaign of Redfin's mobile app (``Redfin Real Estate'') advertised a \installonly offer and a \appusage offer during the second quarter of 2019 as observed in our data collection.
As another example, the quarterly report of IGG, a publicly traded mobile game company, praised its app ``Lords Mobile'' for staying in the top 5 of the top charts in 54 countries during the second quarter of 2019~\cite{q2IGG}.
We observed a total of 18 incentivized \appusage offers for this app over the duration of 3 months; these incentivized install campaigns could potentially have helped the app maintain its top ranks.

These public companies could be purchasing incentivized installs to artificially inflate their quarterly reports.
Nevertheless, it is also possible that these companies are being defrauded by \incentservices, as we discuss next in Section \ref{sec: discussion}.

\noindent \textbf{Summary.}
Overall, our results show a correlation between advertising apps on \incentservices and developers of advertised apps raising funding from investors.
This suggests that inflation of user engagement metrics through incentivized install campaigns may have misled investors.


\presec \section{Discussion} \label{sec: discussion} \postsec

Our analysis characterizes for the first time (and at scale) an unexplored part of the mobile ecosystem, and of alternative mobile app monetization strategies. We reveal a correlation between the apps participating in incentivized install campaigns and (1) their increase in install counts, (2) their appearance on \googlePlay's top charts, and (3) in some cases, a correlation with successful investor rounds. The ability of such campaigns to boost install counts (and active users) can make apps appear more attractive than they actually are by manipulating \googlePlay's rankings. This practice, prohibited on the \iosstore, could be considered as potentially deceptive to Android users and investors alike, as it could give an unfair advantage to apps that are willing to pay users for installs over those that rely on organic growth strategies.

Yet, this does not imply that apps appearing on incentivized install services are willingly engaging in such activities. It is plausible that some app developers are themselves being deceived by third-party marketing organizations offering mechanisms to attract new users or improve their brand value. 
For example, marketers might attempt to arbitrage advertising campaigns by charging for more expensive non-incentivized installs, yet delivering lower-quality incentivized installs \cite{incentInstallMislabelling}. 
In fact, during our measurements, we observed campaigns involving apps published by well-known brands such as Amazon, Apple, and LinkedIn.
As we discuss next, we disclose our findings to app developers.

\subsection{Responsible Disclosure}

We contacted the developers of popular apps advertised on vetted and unvetted \incentservices for responsible disclosure. 
%
We contacted only 136 popular apps, each  with 5M+ installs, out of 922 total apps since this was a manual process using the contact email address from their \googlePlay profile.
%
%
We informed the developers that their apps were advertised as part of an incentivized install campaign.
%
At the time of writing, we have received responses from three developers, all of whom were unaware of their apps participating in such campaigns. 
%
%
%
They also indicated that they are being defrauded.
For example, one developer indicated that they contracted multiple external marketing organizations to acquire non-incentivized installs.
%
It is possible that one of these marketing organizations advertised the developer's app on \incentservices during the time of our data collection.
%

We also disclosed our findings to Google since our results indicate that many of the advertised apps and \incentservices are potentially violating \googlePlay's policy.
Other than the receipt of acknowledgement, we have so far not received any other feedback from Google either corroborating or refuting our findings.

\presub \subsection{\googlePlay's Policy and Enforcement} \postsub
\googlePlay's policy \cite{incentInstallPolicyGoogle} states that \emph{``Developers must not attempt to manipulate the placement of any apps in Google Play.''} The presence and behavior of incentivized install platforms seem to directly contradict this requirement. In fact, some incentivized install platforms like \rankapp (see Figure \ref{fig: rankapp}) are operational on \googlePlay despite openly advertising services that seem to run afoul of these policies.

Our successful purchase of incentivized installs in Section \ref{sec: activemeasurement} suggested that \googlePlay's detection and filtering systems \cite{spamInstallsGoogleBlog,incentInstallsrankingsGoogleBlog} might fail to effectively enforce their policy against manipulation of app store metrics.
In order to measure potential enforcement, we revisit our collected data and look for a \textit{decrease} in the install counts of advertised apps after the start of their campaign. 
Such a decrease would be an indicator that \googlePlay has identified and removed incentivized installs.\footnote{We acknowledge that if an app already has a large number of installs (\eg 1M), we may not be able to observe the impact of install filtering.}
Overall, we observed no decrease in the install counts of baseline apps or the apps advertised on vetted \incentservices.
For unvetted \incentservices, we do occasionally observe a decrease in install counts during our data collection, but it affects only 2\% of the advertised apps.
One example was the app ``Phonebook - Contacts manager'' advertised on \ayet, whose install count decreased from 1,000 to 500.

Based on these observations, it seems that the effectiveness of enforcement on the \googlePlay is rather limited.
We advocate moving towards a prohibitive stance regarding incentivized installs, similar to the \iosstore \cite{appleRejectAppsPocktgmr}, to impede such manipulation attempts while increasing the trustworthiness of top charts and install counts on the \googlePlay.
Furthermore, our measurement methodology can also help the \googlePlay in improving its existing defenses by detecting apps that benefit from incentivized campaigns.
Specifically, our proposed measurements can provide a ground truth of apps to help train machine learning models in detecting the lockstep behavior of users who perform similar in-app activities to complete the offer \cite{beutelcopycatch13,CaoSynchoTrapCCS14,Jian14CatchSyncKDD}.
To foster follow-up research, we have also publicly shared our crawled data on \url{https://www.github.com/shehrozef/IncentInstalls}.

\presub \subsection{Limitations} \postsub
Due to the limitations of our methodology, we cannot currently explore other aspects of the \incentservice ecosystem:

\begin{itemize}[leftmargin=*]

\item We lack \googlePlay data of apps' profiles and top charts outside of our crawl dates. Having historical and precise metadata for every app -- including their presence in top charts -- would have allowed us to better characterize the use and impact of \incentservices on app store metrics.

\item We lack complete visibility of the whole incentivized install ecosystem. 
We purchased installs for our \honeyapp from three \incentservices. We did not purchase installs from more \incentservices due to budget limitations. 
In our in-the-wild measurement, our instrumentation of 8 affiliate apps allowed us to monitor 7 \incentservices. We did not instrument more affiliate apps since it entails significant engineering effort.

\item Comparing results of purchasing incentivized installs of our \honeyapp with non-incentivized installs would help us better understand and distinguish the install quality and \googlePlay's existing enforcement. We did not purchase non-incentivized installs due to budget limitations.

\item We cannot estimate the amount of money spent by an incentivized install campaign of a mobile app because we lack data on the number of offers completed by users. We are only able to estimate the payout to complete an advertised offer that is observable in the offer wall's traffic data.

\item 
Our work focuses on studying IIP activity through
affiliate apps. While other methods to distribute offers exist
(such as micro service platforms), they fall out of the scope of this paper and require developing different methodologies that could be further studied in future work.

\item There are potential grey areas allowing developers running incentivized install campaigns to increase revenue that we did not explore in our study.
The completion of some in-app tasks may require users to purchase gaming credits, virtual currencies, or premium subscriptions, which are not mentioned in the offer description.
We would not be able to identify these tasks since our existing categorization of in-app tasks relies on reading offer descriptions.
Understanding these behaviors would require more extensive and comprehensive use of advertised apps (\eg automating interaction with advertised apps).

\end{itemize}


\presec \section{Related Work}\label{sec:relatedwork} \postsec
To the best of our knowledge, we are the first to study the characteristics of incentivized install services for mobile apps, and the possibilities they offer for manipulating in-app metrics such as user engagement or revenue. Prior work addressed similar, but distinct, areas of pay-per-install (PPI) services for distributing malicious or unwanted desktop apps, and reputation manipulation for online social networks and mobile app stores as discussed next.

\noindent \textbf{App Install Campaigns on Desktop.}
In contrast to the mobile app install campaigns we observe in this paper, desktop app install campaigns appear to be predominantly used by attackers and malicious actors to disseminate malware and potentially unwanted software (PUP) through deceptive practices and compromised hosts~\cite{caballero2011ppiUSENIX,thomas16ppiUSENIX,kotzias16ppiUSENIX}. 
Caballero \etal found that such campaigns installed malicious apps from 12 out of 20 prevalent malware families on compromised host machines~\cite{caballero2011ppiUSENIX}. 
Kotzias \etal studied the presence of potentially unwanted software distributed through app install campaigns and detected at least one sample of potentially unwanted software on 54\% of 9 million hosts~\cite{kotzias16ppiUSENIX}. These services rely on deceptive practices such as drive-by downloads~\cite{moshchuk2006crawler} or default opt-in to install
potentially unwanted software.

The ecosystem of incentivized mobile app installs differs from desktop app installs in at least two ways.
First, users participating in mobile app installs consciously install an app to receive a reward, whereas desktop app installs typically occur without the user's awareness, such as through deceptive practices, or covertly on compromised hosts.
Second, the apps distributed through the install campaigns in desktop settings tend to be PUP or malware.
In contrast, the install campaigns in mobile advertise predominantly benign apps, presumably due to the presence of gatekeepers such as Google's Play Protect program to prevent the distribution of malware. Furthermore, incentivized install providers seem to base their business on artificially increasing the reputation of the apps with the goal of driving future business opportunities and revenues.

\noindent \textbf{Reputation Manipulation.}
Prior work has studied reputation manipulation 
on online social networks to artificially inflate 
social metrics such as likes, followers, or retweets in Facebook, Twitter, and Instagram~\cite{Stringhini13FollowGreenIMC,Viswanath2014AnomalousUSENIX,Song15CrowdTargetCCS,FarooqiOauthAbuse17IMC,Farooqi17CharacterizingBlackHateCrime,ikram17likeFarmsTOPS,DeKovenFollowFootstepInsta2018IMC,Weerasinghe20PodWWW}.
Similar to install counts of mobile apps, manipulating these metrics can increase the perceived value of a social media profile.
Incentivized mobile app installs differ from social network reputation manipulation in the ecosystem of key actors and the methods they employ, such as detailed task ``offers'' distributed to crowd workers through a network of affiliate apps instead of using automation.
These tasks often go beyond the mere installation of apps, requesting for example that workers reach a certain level in a game, or make in-app purchases before they can get paid.

\noindent \textbf{Manipulating App Store Charts.} 
A different line of prior work focused on the detection of app ranking manipulation in marketplaces according to publicly available metrics such as reviews, ratings, and ranking patterns of apps in app stores' top 
charts~\cite{zhu2013ranking,chen17collusiveRankingASIACCS,rahman17fairPlaySIAM,rahman18searchfraudHT}.
For example, Zhu~\etal~\cite{zhu2013ranking} built a statistical behavioral model of historical ranking patterns and the rating score of apps to detect manipulation.
In contrast, our work detects actual apps engaging in manipulation by monitoring ``offer walls'' rather than making statistical inferences.
Our methodology reveals tasks aiming to manipulate app store metrics that are not publicly visible and not targeted at improving the app's standing in the app store, such as inflating in-app revenue or user engagement.
Prior work has also interviewed fraudsters selling fake installs, reviews, and ratings in \googlePlay to reveal their operational characteristics~\cite{rahman19artCraftCCS,hernandez18fraudDeanonymizeCCS}.
While we do not directly interact with the people behind the incentivized install platforms, our work sheds light on the ecosystem at scale, and provides new insights into how these services operate, as well as how app developers may stand to benefit from using these services.

\presec \section{Conclusion} \postsec
In this paper, we provided an in-depth analysis of the ecosystem of incentivized installs on \googlePlay by purchasing installs for our \honeyapp and monitoring \incentservices. 
We find that most of the users are potentially crowd workers installing apps advertised on \incentservices to earn money.
Our results also showed that \incentservices can have negative effects on the app store and beyond (\eg deceitfully raise funding and erode consumer trust) due to their ability to manipulate user engagement metrics.
Finally, our results demonstrate a likely lack of enforcement from \googlePlay to detect the potential violations of their policies by incentivized installs.

Our methodology to investigate IIPs could be deployed to provide a semi-automated monitoring service of \incentservices. This could be leveraged by the \googlePlay to act on potential violations of its terms and services. It could also be leveraged by apps conducting non-incentivized marketing campaigns to detect if a third-party is deceiving them and delivering incentivized installs, and by investors during due diligence investigations. We believe this would help increase the trustworthiness of user engagement metrics in \googlePlay.

\section*{Acknowledgement}
We would like to thank our shepherd, Maria Papadopouli, and the anonymous reviewers for their useful feedback on this paper.
We would also like to thank Sojhal Ismail, Abubakar Aziz, and Muzammil Hussain for their help with data collection. 
This work is supported in part by the National Science Foundation (under grant numbers 1564329, 1715152, 1814816, and 1954224), the European Union's Horizon 2020 Innovation Action program (grant agreement number 786741, SMOOTH Project), the Spanish Ministry of Science and Innovation (grant agreement number PID2019-111429RB-C22, ODIO project), and by unrestricted gifts from Facebook and Google.

\balance
\bibliographystyle{ACM-Reference-Format}
\bibliography{shehroze}


\begin{thebibliography}{68}


\ifx \showCODEN    \undefined \def \showCODEN     #1{\unskip}     \fi
\ifx \showDOI      \undefined \def \showDOI       #1{#1}\fi
\ifx \showISBNx    \undefined \def \showISBNx     #1{\unskip}     \fi
\ifx \showISBNxiii \undefined \def \showISBNxiii  #1{\unskip}     \fi
\ifx \showISSN     \undefined \def \showISSN      #1{\unskip}     \fi
\ifx \showLCCN     \undefined \def \showLCCN      #1{\unskip}     \fi
\ifx \shownote     \undefined \def \shownote      #1{#1}          \fi
\ifx \showarticletitle \undefined \def \showarticletitle #1{#1}   \fi
\ifx \showURL      \undefined \def \showURL       {\relax}        \fi
\providecommand\bibfield[2]{#2}
\providecommand\bibinfo[2]{#2}
\providecommand\natexlab[1]{#1}
\providecommand\showeprint[2][]{arXiv:#2}

\bibitem[\protect\citeauthoryear{??}{q2I}{[n.d.]}]%
        {q2IGG}
 \bibinfo{year}{[n.d.]}\natexlab{}.
\newblock \bibinfo{title}{{2019 Interim Report iGG}}.
\newblock
  \bibinfo{howpublished}{{\url{http://iis.quamnet.com/media/IRAnnouncement/799/EN_US/003588684-0.PDF}}}.
\newblock


\bibitem[\protect\citeauthoryear{??}{iOS}{[n.d.]}]%
        {iOSappStorePolicy}
 \bibinfo{year}{[n.d.]}\natexlab{}.
\newblock \bibinfo{title}{{App Store Review Guidelines}}.
\newblock
  \bibinfo{howpublished}{{https://developer.apple.com/app-store/review/guidelines/}}.
\newblock


\bibitem[\protect\citeauthoryear{??}{chi}{[n.d.]}]%
        {chiSqExample}
 \bibinfo{year}{[n.d.]}\natexlab{}.
\newblock \bibinfo{title}{{Chi-Square Test for Independence}}.
\newblock
  \bibinfo{howpublished}{{\url{https://stattrek.com/chi-square-test/independence.aspx}}}.
\newblock


\bibitem[\protect\citeauthoryear{??}{ios}{[n.d.]}]%
        {iosBusinessModel}
 \bibinfo{year}{[n.d.]}\natexlab{}.
\newblock \bibinfo{title}{{Choosing a Business Model}}.
\newblock
  \bibinfo{howpublished}{{\url{https://developer.apple.com/app-store/business-models/}}}.
\newblock


\bibitem[\protect\citeauthoryear{??}{inv}{[n.d.]}]%
        {investorInstallsStartups}
 \bibinfo{year}{[n.d.]}\natexlab{}.
\newblock \bibinfo{title}{{Do angel investors look for a certain number of
  active users when investing in an app startup that has recently launched,
  5,000-10,000+?}}
\newblock
  \bibinfo{howpublished}{{\url{https://www.startups.com/community/questions/1841/do-angel-investors-look-for-a-certain-number-of-active-users-when-investing-in}}}.
\newblock


\bibitem[\protect\citeauthoryear{??}{gpl}{[n.d.]a}]%
        {gplayBusinessModel}
 \bibinfo{year}{[n.d.]}\natexlab{a}.
\newblock \bibinfo{title}{{Earn more revenue with the right monetization
  options}}.
\newblock
  \bibinfo{howpublished}{{\url{https://developer.android.com/distribute/best-practices/earn/monetization-options}}}.
\newblock


\bibitem[\protect\citeauthoryear{??}{fyb}{[n.d.]}]%
        {fyberOwler}
 \bibinfo{year}{[n.d.]}\natexlab{}.
\newblock \bibinfo{title}{{Fyber Competitors, Revenues and Employees}}.
\newblock \bibinfo{howpublished}{{\url{https://www.owler.com/company/fyber}}}.
\newblock


\bibitem[\protect\citeauthoryear{??}{inc}{[n.d.]a}]%
        {incentInstallMislabelling}
 \bibinfo{year}{[n.d.]}\natexlab{a}.
\newblock \bibinfo{title}{{INCENTIVIZED INSTALL MISLABELING}}.
\newblock
  \bibinfo{howpublished}{{\url{https://appsflyer.com/glossary/incentivized-install-mislabeling/}}}.
\newblock


\bibitem[\protect\citeauthoryear{??}{app}{[n.d.]}]%
        {appInstallCost}
 \bibinfo{year}{[n.d.]}\natexlab{}.
\newblock \bibinfo{title}{{Mobile App User Acquistion Cost (2019)}}.
\newblock
  \bibinfo{howpublished}{{\url{https://www.statista.com/statistics/185736/mobile-app-average-user-acquisition-cost/}}}.
\newblock


\bibitem[\protect\citeauthoryear{??}{roo}{[n.d.]}]%
        {rootBeer}
 \bibinfo{year}{[n.d.]}\natexlab{}.
\newblock \bibinfo{title}{{RootBeer}}.
\newblock \bibinfo{howpublished}{{\url{https://github.com/scottyab/rootbeer}}}.
\newblock


\bibitem[\protect\citeauthoryear{??}{inc}{[n.d.]b}]%
        {incentInstallPolicyGoogle}
 \bibinfo{year}{[n.d.]}\natexlab{b}.
\newblock \bibinfo{title}{{User Ratings, Reviews, and Installs}}.
\newblock
  \bibinfo{howpublished}{{https://play.google.com/about/storelisting-promotional/ratings-reviews-installs/}}.
\newblock


\bibitem[\protect\citeauthoryear{??}{gpl}{[n.d.]b}]%
        {gplayDevStats}
 \bibinfo{year}{[n.d.]}\natexlab{b}.
\newblock \bibinfo{title}{{View app statistics}}.
\newblock
  \bibinfo{howpublished}{{https://support.google.com/googleplay/android-developer/answer/139628}}.
\newblock


\bibitem[\protect\citeauthoryear{??}{gro}{2018}]%
        {growthVsProfit}
 \bibinfo{year}{2018}\natexlab{}.
\newblock \bibinfo{title}{{A Healthy Tension Between Growth and Profit is a
  Common Phenomenon For Any Startup}}.
\newblock
  \bibinfo{howpublished}{{\url{https://theaccountancycloud.com/resources/blogs/a-healthy-tension-between-growth-and-profit-is-a-common-phenomenon-for-any-startup}}}.
\newblock


\bibitem[\protect\citeauthoryear{??}{fyb}{2019}]%
        {fyberReport}
 \bibinfo{year}{2019}\natexlab{}.
\newblock \bibinfo{title}{{Fyber N.V. published results for Q1 2019}}.
\newblock
  \bibinfo{howpublished}{{\url{https://investors.fyber.com/assets/media/fyber-nv-q1-2019-results-announcement-english.pdf}}}.
\newblock


\bibitem[\protect\citeauthoryear{{Appium}}{{Appium}}{[n.d.]}]%
        {appium}
\bibfield{author}{\bibinfo{person}{{Appium}}.}
  \bibinfo{year}{[n.d.]}\natexlab{}.
\newblock \bibinfo{title}{{Automation for apps}}.
\newblock \bibinfo{howpublished}{{\url{appium.io}}}.
\newblock


\bibitem[\protect\citeauthoryear{{Artyom Dogtiev}}{{Artyom Dogtiev}}{2020}]%
        {appAdPlatforms}
\bibfield{author}{\bibinfo{person}{{Artyom Dogtiev}}.}
  \bibinfo{year}{2020}\natexlab{}.
\newblock \bibinfo{title}{{Top App Install Ad Platforms (2019)}}.
\newblock
  \bibinfo{howpublished}{{\url{businessofapps.com/guide/app-install-ad-networks/}}}.
\newblock


\bibitem[\protect\citeauthoryear{Bailey, Dittrich, Kenneally, and
  Maughan}{Bailey et~al\mbox{.}}{2012}]%
        {bailey12MenloSP}
\bibfield{author}{\bibinfo{person}{Michael Bailey}, \bibinfo{person}{David
  Dittrich}, \bibinfo{person}{Erin Kenneally}, {and} \bibinfo{person}{Doug
  Maughan}.} \bibinfo{year}{2012}\natexlab{}.
\newblock \showarticletitle{{The Menlo report}}.
\newblock \bibinfo{journal}{\emph{IEEE Security \& Privacy 10 (2)}}
  (\bibinfo{year}{2012}), \bibinfo{pages}{71--75}.
\newblock


\bibitem[\protect\citeauthoryear{Beutel, Xu, Wenkatesan, Chirstopher, and
  Christos}{Beutel et~al\mbox{.}}{2013}]%
        {beutelcopycatch13}
\bibfield{author}{\bibinfo{person}{Alex Beutel}, \bibinfo{person}{Wanhong Xu},
  \bibinfo{person}{Wenkatesan}, \bibinfo{person}{Chirstopher}, {and}
  \bibinfo{person}{Christos}.} \bibinfo{year}{2013}\natexlab{}.
\newblock \showarticletitle{{CopyCatch: Stopping Group Attacks by Spotting
  Lockstep Behavior in Social Networks}}. In \bibinfo{booktitle}{\emph{WWW}}.
\newblock


\bibitem[\protect\citeauthoryear{Caballero, Grier, Kreibich, and
  Paxson}{Caballero et~al\mbox{.}}{2011}]%
        {caballero2011ppiUSENIX}
\bibfield{author}{\bibinfo{person}{Juan Caballero}, \bibinfo{person}{Chris
  Grier}, \bibinfo{person}{Christian Kreibich}, {and} \bibinfo{person}{Vern
  Paxson}.} \bibinfo{year}{2011}\natexlab{}.
\newblock \showarticletitle{Measuring pay-per-install: The commoditization of
  malware distribution}. In \bibinfo{booktitle}{\emph{{USENIX Security
  Symposium}}}.
\newblock


\bibitem[\protect\citeauthoryear{Cao, Yang, Yu, and Palow}{Cao
  et~al\mbox{.}}{2014}]%
        {CaoSynchoTrapCCS14}
\bibfield{author}{\bibinfo{person}{Qiang Cao}, \bibinfo{person}{Xiaowei Yang},
  \bibinfo{person}{Jieqi Yu}, {and} \bibinfo{person}{Christopher Palow}.}
  \bibinfo{year}{2014}\natexlab{}.
\newblock \showarticletitle{{Uncovering Large Groups of Active Malicious
  Accounts in Online Social Networks}}. In \bibinfo{booktitle}{\emph{ACM CSS}}.
\newblock


\bibitem[\protect\citeauthoryear{{Caroline Banton}}{{Caroline Banton}}{2020}]%
        {defPublicComp}
\bibfield{author}{\bibinfo{person}{{Caroline Banton}}.}
  \bibinfo{year}{2020}\natexlab{}.
\newblock \bibinfo{title}{{Public Company}}.
\newblock
  \bibinfo{howpublished}{{\url{https://www.investopedia.com/terms/p/publiccompany.asp}}}.
\newblock


\bibitem[\protect\citeauthoryear{Chapple}{Chapple}{2019}]%
        {appleRejectAppsPocktgmr}
\bibfield{author}{\bibinfo{person}{Craig Chapple}.}
  \bibinfo{year}{2019}\natexlab{}.
\newblock \bibinfo{title}{{Devs lose out on thousands of dollars as Apple
  cracks down on offerwall ads}}.
\newblock
  \bibinfo{howpublished}{{https://www.pocketgamer.biz/news/70610/devs-lose-thousands-of-dollars-as-apple-cracks-down-on-offerwall-ads/}}.
\newblock


\bibitem[\protect\citeauthoryear{Chen, He, Zhu, and Yang}{Chen
  et~al\mbox{.}}{2017}]%
        {chen17collusiveRankingASIACCS}
\bibfield{author}{\bibinfo{person}{Hao Chen}, \bibinfo{person}{Daojing He},
  \bibinfo{person}{Sencun Zhu}, {and} \bibinfo{person}{Jingshun Yang}.}
  \bibinfo{year}{2017}\natexlab{}.
\newblock \showarticletitle{Toward detecting collusive ranking manipulation
  attackers in mobile app markets}. In \bibinfo{booktitle}{\emph{ACM ASIACCS}}.
\newblock


\bibitem[\protect\citeauthoryear{Crunchbase}{Crunchbase}{[n.d.]}]%
        {crunchbase}
\bibfield{author}{\bibinfo{person}{Crunchbase}.}
  \bibinfo{year}{[n.d.]}\natexlab{}.
\newblock \bibinfo{title}{{Crunchbase}}.
\newblock \bibinfo{howpublished}{{www.crunchbase.com}}.
\newblock


\bibitem[\protect\citeauthoryear{DeKoven, Pottinger, Savage, Voelker, and
  Leontiadis}{DeKoven et~al\mbox{.}}{2018}]%
        {DeKovenFollowFootstepInsta2018IMC}
\bibfield{author}{\bibinfo{person}{Louis DeKoven}, \bibinfo{person}{Trevor
  Pottinger}, \bibinfo{person}{Stefan Savage}, \bibinfo{person}{Geoffrey
  Voelker}, {and} \bibinfo{person}{Nektarios Leontiadis}.}
  \bibinfo{year}{2018}\natexlab{}.
\newblock \showarticletitle{{Following Their Footsteps: Characterizing Account
  Automation Abuse and Defenses}}. In \bibinfo{booktitle}{\emph{ACM Internet
  Measurement Conference (IMC)}}.
\newblock


\bibitem[\protect\citeauthoryear{Farooqi, Ikram, Cristofaro, Friedman, Jourjon,
  Kaafar, Shafiq, and Zaffar}{Farooqi et~al\mbox{.}}{2017a}]%
        {Farooqi17CharacterizingBlackHateCrime}
\bibfield{author}{\bibinfo{person}{Shehroze Farooqi}, \bibinfo{person}{Muhammad
  Ikram}, \bibinfo{person}{Emiliano~De Cristofaro}, \bibinfo{person}{Arik
  Friedman}, \bibinfo{person}{Guillaume Jourjon}, \bibinfo{person}{Mohamed
  Kaafar}, \bibinfo{person}{Zubair Shafiq}, {and} \bibinfo{person}{Fareed
  Zaffar}.} \bibinfo{year}{2017}\natexlab{a}.
\newblock \showarticletitle{{Characterizing Key Stakeholders in an Online
  Black-Hat Marketplace}}. In \bibinfo{booktitle}{\emph{IEEE/APWG Symposium on
  Electronic Crime Research (eCrime)}}.
\newblock


\bibitem[\protect\citeauthoryear{Farooqi, Zaffar, Leontiadis, and
  Shafiq}{Farooqi et~al\mbox{.}}{2017b}]%
        {FarooqiOauthAbuse17IMC}
\bibfield{author}{\bibinfo{person}{Shehroze Farooqi}, \bibinfo{person}{Fareed
  Zaffar}, \bibinfo{person}{Nektarios Leontiadis}, {and}
  \bibinfo{person}{Zubair Shafiq}.} \bibinfo{year}{2017}\natexlab{b}.
\newblock \showarticletitle{{Measuring and Mitigating OAuth Access Token Abuse
  by Collusion Networks}}. In \bibinfo{booktitle}{\emph{ACM Internet
  Measurement Conference (IMC)}}.
\newblock


\bibitem[\protect\citeauthoryear{{Geri Terzo}}{{Geri Terzo}}{2020}]%
        {stockPriceEarning}
\bibfield{author}{\bibinfo{person}{{Geri Terzo}}.}
  \bibinfo{year}{2020}\natexlab{}.
\newblock \bibinfo{title}{{The Impact of Earnings Announcements on Stock
  Prices}}.
\newblock
  \bibinfo{howpublished}{\url{https://finance.zacks.com/impact-earnings-announcements-stock-prices-4265.html}}.
\newblock


\bibitem[\protect\citeauthoryear{{Gurdeep Singh}}{{Gurdeep Singh}}{2018}]%
        {apple_ios_users}
\bibfield{author}{\bibinfo{person}{{Gurdeep Singh}}.}
  \bibinfo{year}{2018}\natexlab{}.
\newblock \bibinfo{title}{{29 App Store Stats 2018-19 that Proves Apple’s
  Uprising Growth Trend}}.
\newblock
  \bibinfo{howpublished}{{\url{https://appinventiv.com/blog/apple-app-store-statistics/}}}.
\newblock


\bibitem[\protect\citeauthoryear{Haddadi}{Haddadi}{2010}]%
        {haddadi10ClickFraud}
\bibfield{author}{\bibinfo{person}{Hamed Haddadi}.}
  \bibinfo{year}{2010}\natexlab{}.
\newblock \showarticletitle{{Fighting online click-fraud using bluff ads}}.
\newblock \bibinfo{journal}{\emph{ACM SIGCOMM Computer Communication Review}}
  \bibinfo{volume}{40}, \bibinfo{number}{2} (\bibinfo{year}{2010}),
  \bibinfo{pages}{21--25}.
\newblock


\bibitem[\protect\citeauthoryear{Hernandez, Rahman, Recabarren, and
  Carbunar}{Hernandez et~al\mbox{.}}{2018}]%
        {hernandez18fraudDeanonymizeCCS}
\bibfield{author}{\bibinfo{person}{Nestor Hernandez}, \bibinfo{person}{Mizanur
  Rahman}, \bibinfo{person}{Ruben Recabarren}, {and} \bibinfo{person}{Bogdan
  Carbunar}.} \bibinfo{year}{2018}\natexlab{}.
\newblock \showarticletitle{Fraud de-anonymization for fun and profit}. In
  \bibinfo{booktitle}{\emph{ACM CCS}}.
\newblock


\bibitem[\protect\citeauthoryear{Ikram, Onwuzurike, Farooqi, Cristofaro,
  Friedman, Jourjon, Kaafar, and Shafiq}{Ikram et~al\mbox{.}}{2017}]%
        {ikram17likeFarmsTOPS}
\bibfield{author}{\bibinfo{person}{Muhammad Ikram}, \bibinfo{person}{Lucky
  Onwuzurike}, \bibinfo{person}{Shehroze Farooqi}, \bibinfo{person}{Emiliano~De
  Cristofaro}, \bibinfo{person}{Arik Friedman}, \bibinfo{person}{Guillaume
  Jourjon}, \bibinfo{person}{Mohammed~Ali Kaafar}, {and}
  \bibinfo{person}{Zubair Shafiq}.} \bibinfo{year}{2017}\natexlab{}.
\newblock \showarticletitle{{Measuring, Characterizing, and Detecting Facebook
  Like Farms}}.
\newblock \bibinfo{journal}{\emph{ACM Transactions on Privacy and Security}}
  \bibinfo{volume}{20}, \bibinfo{number}{4} (\bibinfo{year}{2017}),
  \bibinfo{pages}{1--28}.
\newblock


\bibitem[\protect\citeauthoryear{Jian, Cui, Beutel, Faloutsos, and Yang}{Jian
  et~al\mbox{.}}{2014}]%
        {Jian14CatchSyncKDD}
\bibfield{author}{\bibinfo{person}{Meng Jian}, \bibinfo{person}{Peng Cui},
  \bibinfo{person}{Alex Beutel}, \bibinfo{person}{Christos Faloutsos}, {and}
  \bibinfo{person}{Shiqiang Yang}.} \bibinfo{year}{2014}\natexlab{}.
\newblock \showarticletitle{{CatchSync: catching synchronized behavior in large
  directed graphs}}. In \bibinfo{booktitle}{\emph{ACM KDD}}.
\newblock


\bibitem[\protect\citeauthoryear{{Joseph Cox}}{{Joseph Cox}}{2019}]%
        {deviceFarmVice}
\bibfield{author}{\bibinfo{person}{{Joseph Cox}}.}
  \bibinfo{year}{2019}\natexlab{}.
\newblock \bibinfo{title}{{America’s DIY Phone Farmers}}.
\newblock
  \bibinfo{howpublished}{{\url{https://www.vice.com/en_us/article/d3naek/how-to-make-a-phone-farm}},Vice}.
\newblock


\bibitem[\protect\citeauthoryear{Kim}{Kim}{2017}]%
        {investorGrowthForbes}
\bibfield{author}{\bibinfo{person}{Jay Kim}.} \bibinfo{year}{2017}\natexlab{}.
\newblock \bibinfo{title}{{How To Catch The Attention Of One Of The Most Active
  Startup Investors In The World}}.
\newblock
  \bibinfo{howpublished}{{\url{https://www.forbes.com/sites/kimjay/2017/11/02/how-to-catch-the-attention-of-one-of-the-most-active-startup-investors-in-the-world/}}}.
\newblock


\bibitem[\protect\citeauthoryear{Kotzias, Bilge, and Caballero}{Kotzias
  et~al\mbox{.}}{2016}]%
        {kotzias16ppiUSENIX}
\bibfield{author}{\bibinfo{person}{Platon Kotzias}, \bibinfo{person}{Leyla
  Bilge}, {and} \bibinfo{person}{Juan Caballero}.}
  \bibinfo{year}{2016}\natexlab{}.
\newblock \showarticletitle{Measuring PUP Prevalence and PUP Distribution
  through Pay-Per-Install Services}. In \bibinfo{booktitle}{\emph{USENIX
  Security Symposium}}.
\newblock


\bibitem[\protect\citeauthoryear{Ma, Wang, Guo, and Chen}{Ma
  et~al\mbox{.}}{2016}]%
        {ma2016libradar}
\bibfield{author}{\bibinfo{person}{Ziang Ma}, \bibinfo{person}{Haoyu Wang},
  \bibinfo{person}{Yao Guo}, {and} \bibinfo{person}{Xiangqun Chen}.}
  \bibinfo{year}{2016}\natexlab{}.
\newblock \showarticletitle{{LibRadar: fast and accurate detection of
  third-party libraries in Android apps}}. In
  \bibinfo{booktitle}{\emph{International Conference on Software Engineering
  (ICSE) Companion}}.
\newblock


\bibitem[\protect\citeauthoryear{Mashable}{Mashable}{2015}]%
        {play_billion}
\bibfield{author}{\bibinfo{person}{Mashable}.} \bibinfo{year}{2015}\natexlab{}.
\newblock \bibinfo{title}{{Google Play just hit a major milestone}}.
\newblock
  \bibinfo{howpublished}{{https://mashable.com/2015/09/29/google-play-1-billion-users/}}.
\newblock


\bibitem[\protect\citeauthoryear{McHugh}{McHugh}{2013}]%
        {chiSquareTest}
\bibfield{author}{\bibinfo{person}{Mary McHugh}.}
  \bibinfo{year}{2013}\natexlab{}.
\newblock \showarticletitle{The chi-square test of independence}.
\newblock \bibinfo{journal}{\emph{Biochemia medica: Biochemia medica}}
  (\bibinfo{year}{2013}).
\newblock


\bibitem[\protect\citeauthoryear{{mitmproxy}}{{mitmproxy}}{[n.d.]}]%
        {mitm}
\bibfield{author}{\bibinfo{person}{{mitmproxy}}.}
  \bibinfo{year}{[n.d.]}\natexlab{}.
\newblock \bibinfo{title}{{Homepage}}.
\newblock \bibinfo{howpublished}{{\url{mitmproxy.org}}}.
\newblock


\bibitem[\protect\citeauthoryear{Moshchuk, Bragin, Gribble, and Levy}{Moshchuk
  et~al\mbox{.}}{2006}]%
        {moshchuk2006crawler}
\bibfield{author}{\bibinfo{person}{Alexander Moshchuk}, \bibinfo{person}{Tanya
  Bragin}, \bibinfo{person}{Steven~D Gribble}, {and} \bibinfo{person}{Henry~M
  Levy}.} \bibinfo{year}{2006}\natexlab{}.
\newblock \showarticletitle{A Crawler-based Study of Spyware in the Web}. In
  \bibinfo{booktitle}{\emph{Network and Distributed Systems Security (NDSS)
  Symposium}}.
\newblock


\bibitem[\protect\citeauthoryear{Nagayama}{Nagayama}{2016}]%
        {spamInstallsGoogleBlog}
\bibfield{author}{\bibinfo{person}{Kazushi Nagayama}.}
  \bibinfo{year}{2016}\natexlab{}.
\newblock \bibinfo{title}{{Keeping the Play Store trusted: fighting fraud and
  spam installs}}.
\newblock
  \bibinfo{howpublished}{{https://android-developers.googleblog.com/2016/10/keeping-the-play-store-trusted-fighting-fraud-and-spam-installs.html},
  Android Developers Blog}.
\newblock


\bibitem[\protect\citeauthoryear{Nagayama}{Nagayama}{2017}]%
        {incentInstallsrankingsGoogleBlog}
\bibfield{author}{\bibinfo{person}{Kazushi Nagayama}.}
  \bibinfo{year}{2017}\natexlab{}.
\newblock \bibinfo{title}{{Google Play’s policy on incentivized ratings,
  reviews, and installs}}.
\newblock
  \bibinfo{howpublished}{{https://android-developers.googleblog.com/2017/06/google-plays-policy-on-incentivized.html},
  Android Developers Blog}.
\newblock


\bibitem[\protect\citeauthoryear{{Paul Bankhead}}{{Paul Bankhead}}{2018}]%
        {engagementVisibilityGplay}
\bibfield{author}{\bibinfo{person}{{Paul Bankhead}}.}
  \bibinfo{year}{2018}\natexlab{}.
\newblock \bibinfo{title}{{Improving discovery of quality apps and games on the
  Play Store}}.
\newblock
  \bibinfo{howpublished}{{\url{http://ndrdnws.blogspot.com/2018/06/improving-discovery-of-quality-apps-and.html}}}.
\newblock


\bibitem[\protect\citeauthoryear{Pearson}{Pearson}{2015}]%
        {eyeballAds}
\bibfield{author}{\bibinfo{person}{Phil Pearson}.}
  \bibinfo{year}{2015}\natexlab{}.
\newblock \bibinfo{title}{Eyeballs vs Engagement: the Struggle of Digital
  Advertising}.
\newblock
  \bibinfo{howpublished}{\url{https://medium.com/@philgpearson/eyeballs-vs-engagement-the-struggle-of-digital-advertising-301ffd61c773},Medium}.
\newblock


\bibitem[\protect\citeauthoryear{Prez}{Prez}{2019}]%
        {dashlane-1}
\bibfield{author}{\bibinfo{person}{Sarah Prez}.}
  \bibinfo{year}{2019}\natexlab{}.
\newblock \bibinfo{title}{{Password manager Dashlane closes on \$30M, adds
  former Spotify CMO to board}}.
\newblock
  \bibinfo{howpublished}{{https://techcrunch.com/2019/04/12/password-manager-dashlane-closes-on-30m-adds-former-spotify-cmo-to-board/}}.
\newblock


\bibitem[\protect\citeauthoryear{Rahman, Hernandez, Carbunar, and Chau}{Rahman
  et~al\mbox{.}}{2018}]%
        {rahman18searchfraudHT}
\bibfield{author}{\bibinfo{person}{Mizanur Rahman}, \bibinfo{person}{Nestor
  Hernandez}, \bibinfo{person}{Bogdan Carbunar}, {and}
  \bibinfo{person}{Duen~Horng Chau}.} \bibinfo{year}{2018}\natexlab{}.
\newblock \showarticletitle{Search Rank Fraud De-Anonymization in Online
  Systems}. In \bibinfo{booktitle}{\emph{ACM Hypertext and Social Media (HT)}}.
\newblock


\bibitem[\protect\citeauthoryear{Rahman, Hernandez, Recabarren, Ahmed, and
  Carbunar}{Rahman et~al\mbox{.}}{2019}]%
        {rahman19artCraftCCS}
\bibfield{author}{\bibinfo{person}{Mizanur Rahman}, \bibinfo{person}{Nestor
  Hernandez}, \bibinfo{person}{Ruben Recabarren},
  \bibinfo{person}{Syed~Ishtiaque Ahmed}, {and} \bibinfo{person}{Bogdan
  Carbunar}.} \bibinfo{year}{2019}\natexlab{}.
\newblock \showarticletitle{The Art and Craft of Fraudulent App Promotion in
  Google Play}. In \bibinfo{booktitle}{\emph{ACM CCS}}.
\newblock


\bibitem[\protect\citeauthoryear{Rahman, Rahman, Carbunar, and Chau}{Rahman
  et~al\mbox{.}}{2016}]%
        {rahman17fairPlaySIAM}
\bibfield{author}{\bibinfo{person}{Mahmudur Rahman}, \bibinfo{person}{Mizanur
  Rahman}, \bibinfo{person}{Bogdan Carbunar}, {and} \bibinfo{person}{Duen~Horng
  Chau}.} \bibinfo{year}{2016}\natexlab{}.
\newblock \showarticletitle{Fairplay: Fraud and malware detection in google
  play}. In \bibinfo{booktitle}{\emph{International Conference on Data Mining
  SIAM}}.
\newblock


\bibitem[\protect\citeauthoryear{Razaghpanah, Nithyanand, Vallina-Rodriguez,
  Sundaresan, Allman, Kreibich, and Gill}{Razaghpanah et~al\mbox{.}}{2018}]%
        {razaghpanah2018apps}
\bibfield{author}{\bibinfo{person}{Abbas Razaghpanah}, \bibinfo{person}{Rishab
  Nithyanand}, \bibinfo{person}{Narseo Vallina-Rodriguez},
  \bibinfo{person}{Srikanth Sundaresan}, \bibinfo{person}{Mark Allman},
  \bibinfo{person}{Christian Kreibich}, {and} \bibinfo{person}{Phillipa Gill}.}
  \bibinfo{year}{2018}\natexlab{}.
\newblock \showarticletitle{Apps, trackers, privacy, and regulators: A global
  study of the mobile tracking ecosystem}. In \bibinfo{booktitle}{\emph{Network
  and Distributed Systems Security (NDSS) Symposium}}.
\newblock


\bibitem[\protect\citeauthoryear{Razaghpanah, Vallina-Rodriguez, Sundaresan,
  Kreibich, Gill, Allman, and Paxson}{Razaghpanah et~al\mbox{.}}{2015}]%
        {razaghpanah2015haystack}
\bibfield{author}{\bibinfo{person}{Abbas Razaghpanah}, \bibinfo{person}{Narseo
  Vallina-Rodriguez}, \bibinfo{person}{Srikanth Sundaresan},
  \bibinfo{person}{Christian Kreibich}, \bibinfo{person}{Phillipa Gill},
  \bibinfo{person}{Mark Allman}, {and} \bibinfo{person}{Vern Paxson}.}
  \bibinfo{year}{2015}\natexlab{}.
\newblock \showarticletitle{Haystack: In situ mobile traffic analysis in user
  space}.
\newblock \bibinfo{journal}{\emph{arXiv preprint arXiv:1510.01419}}
  (\bibinfo{year}{2015}).
\newblock


\bibitem[\protect\citeauthoryear{{Redfin}}{{Redfin}}{2019}]%
        {redfinQ2}
\bibfield{author}{\bibinfo{person}{{Redfin}}.} \bibinfo{year}{2019}\natexlab{}.
\newblock \bibinfo{title}{{Redfin Second-Quarter 2019 Revenue up 39\%
  Year-over-Year to \$197.8 Million}}.
\newblock
  \bibinfo{howpublished}{{\url{https://www.prnewswire.com/news-releases/redfin-second-quarter-2019-revenue-up-39-year-over-year-to-197-8-million-300895226.html}}}.
\newblock


\bibitem[\protect\citeauthoryear{Reyburn}{Reyburn}{2019}]%
        {fyberiOSIncentInstallBan}
\bibfield{author}{\bibinfo{person}{Scott Reyburn}.}
  \bibinfo{year}{2019}\natexlab{}.
\newblock \bibinfo{title}{{The deprecation of incentivized download offers in
  Offer Wall Edge for iOS, explained}}.
\newblock
  \bibinfo{howpublished}{{https://blog.fyber.com/the-deprecation-of-incentivized-download-offers-in-offer-wall-edge-for-ios-explained/}}.
\newblock


\bibitem[\protect\citeauthoryear{{Saima Salim}}{{Saima Salim}}{2020}]%
        {appstoresRevenuesDigitalInfo}
\bibfield{author}{\bibinfo{person}{{Saima Salim}}.}
  \bibinfo{year}{2020}\natexlab{}.
\newblock \bibinfo{title}{{Apple App Store and Google Play Users Spent over 83
  Billion USD on Mobile Apps in the Last 12 Months, Globally}}.
\newblock
  \bibinfo{howpublished}{{\url{https://www.digitalinformationworld.com/2020/01/global-consumers-spent-over-83-billion-on-mobile-apps-in-the-last-12-months.html}}}.
\newblock


\bibitem[\protect\citeauthoryear{{Sarah Prez}}{{Sarah Prez}}{2012}]%
        {engagementVisibilityGplay_2}
\bibfield{author}{\bibinfo{person}{{Sarah Prez}}.}
  \bibinfo{year}{2012}\natexlab{}.
\newblock \bibinfo{title}{{ASO (App Store Optimization) Is The New SEO, And
  Here’s A Tool To Do It}}.
\newblock
  \bibinfo{howpublished}{{\url{https://techcrunch.com/2012/02/28/aso-app-store-optimization-is-the-new-seo-and-heres-a-tool-to-do-it/}}}.
\newblock


\bibitem[\protect\citeauthoryear{{Shani Rosenfelder}}{{Shani
  Rosenfelder}}{2020}]%
        {appsFlyerAdSpent}
\bibfield{author}{\bibinfo{person}{{Shani Rosenfelder}}.}
  \bibinfo{year}{2020}\natexlab{}.
\newblock \bibinfo{title}{{Global App Install Ad Spend To Double By 2022 to Hit
  118 Billion USD}}.
\newblock
  \bibinfo{howpublished}{{\url{https://www.appsflyer.com/blog/app-install-ad-spend/}}}.
\newblock


\bibitem[\protect\citeauthoryear{Song, Lee, and Kim}{Song
  et~al\mbox{.}}{2015}]%
        {Song15CrowdTargetCCS}
\bibfield{author}{\bibinfo{person}{Jonghyuk Song}, \bibinfo{person}{Sangho
  Lee}, {and} \bibinfo{person}{Jong Kim}.} \bibinfo{year}{2015}\natexlab{}.
\newblock \showarticletitle{{CrowdTarget: Target-based Detection of
  Crowdturfing in Online Social Networks}}. In \bibinfo{booktitle}{\emph{ACM
  CCS}}.
\newblock


\bibitem[\protect\citeauthoryear{Stringhini, Wang, Egele, Kruegel, Vigna,
  Zheng, and Zhao}{Stringhini et~al\mbox{.}}{2013}]%
        {Stringhini13FollowGreenIMC}
\bibfield{author}{\bibinfo{person}{Gianluca Stringhini}, \bibinfo{person}{Gang
  Wang}, \bibinfo{person}{Manuel Egele}, \bibinfo{person}{Christopher Kruegel},
  \bibinfo{person}{Giovanni Vigna}, \bibinfo{person}{Haitao Zheng}, {and}
  \bibinfo{person}{Ben~Y. Zhao}.} \bibinfo{year}{2013}\natexlab{}.
\newblock \showarticletitle{{Follow the Green: Growth and Dynamics in Twitter
  Follower Markets}}. In \bibinfo{booktitle}{\emph{ACM Internet Measurement
  Conference (IMC)}}.
\newblock


\bibitem[\protect\citeauthoryear{Sungwan}{Sungwan}{2019}]%
        {droom}
\bibfield{author}{\bibinfo{person}{Sujata Sungwan}.}
  \bibinfo{year}{2019}\natexlab{}.
\newblock \bibinfo{title}{{Online automobile marketplace Droom gets \$10M in
  Series F from its Singapore-based holding entity}}.
\newblock
  \bibinfo{howpublished}{{https://yourstory.com/2019/06/startup-funding-online-automobile-marketplace-droom}}.
\newblock


\bibitem[\protect\citeauthoryear{Thomas, Crespo, Rasti, Picod, Phillips,
  Decoste, Sharp, Tirelo, Tofigh, Courteau, et~al\mbox{.}}{Thomas
  et~al\mbox{.}}{2016}]%
        {thomas16ppiUSENIX}
\bibfield{author}{\bibinfo{person}{Kurt Thomas}, \bibinfo{person}{Juan
  A.~Elices Crespo}, \bibinfo{person}{Ryan Rasti}, \bibinfo{person}{Jean-Michel
  Picod}, \bibinfo{person}{Cait Phillips}, \bibinfo{person}{Marc-Andre
  Decoste}, \bibinfo{person}{Chris Sharp}, \bibinfo{person}{Fabio Tirelo},
  \bibinfo{person}{Ali Tofigh}, \bibinfo{person}{Marc-Antoine Courteau},
  {et~al\mbox{.}}} \bibinfo{year}{2016}\natexlab{}.
\newblock \showarticletitle{Investigating commercial pay-per-install and the
  distribution of unwanted software}. In \bibinfo{booktitle}{\emph{USENIX
  Security Symposium}}.
\newblock


\bibitem[\protect\citeauthoryear{Viswanath, Bashir, Crovella, Guha, Gummadi,
  Krishnamurthy, and Mislove}{Viswanath et~al\mbox{.}}{2014}]%
        {Viswanath2014AnomalousUSENIX}
\bibfield{author}{\bibinfo{person}{Bimal Viswanath}, \bibinfo{person}{M.~Ahmad
  Bashir}, \bibinfo{person}{Mark Crovella}, \bibinfo{person}{Saikat Guha},
  \bibinfo{person}{Krishna~P. Gummadi}, \bibinfo{person}{Balachander
  Krishnamurthy}, {and} \bibinfo{person}{Alan Mislove}.}
  \bibinfo{year}{2014}\natexlab{}.
\newblock \showarticletitle{{Towards Detecting Anomalous User Behavior in
  Online Social Networks}}. In \bibinfo{booktitle}{\emph{USENIX Security
  Symposium}}.
\newblock


\bibitem[\protect\citeauthoryear{Wagner}{Wagner}{2017}]%
        {investorUser}
\bibfield{author}{\bibinfo{person}{Kurt Wagner}.}
  \bibinfo{year}{2017}\natexlab{}.
\newblock \bibinfo{title}{What kind of apps catch the attention of Silicon
  Valley investors?}
\newblock
  \bibinfo{howpublished}{{\url{https://www.vox.com/2017/11/18/16673982/app-growth-investors-jeremy-liew-snapchat}}}.
\newblock


\bibitem[\protect\citeauthoryear{Weerasinghe, Flanigan, Stein, McCoy, and
  Greenstadt}{Weerasinghe et~al\mbox{.}}{2020}]%
        {Weerasinghe20PodWWW}
\bibfield{author}{\bibinfo{person}{Janith Weerasinghe}, \bibinfo{person}{Bailey
  Flanigan}, \bibinfo{person}{Aviel Stein}, \bibinfo{person}{Damon McCoy},
  {and} \bibinfo{person}{Rachel Greenstadt}.} \bibinfo{year}{2020}\natexlab{}.
\newblock \showarticletitle{{The Pod People: Understanding Manipulation of
  Social Media Popularity via Reciprocity Abuse}}. In
  \bibinfo{booktitle}{\emph{The Web Conference}}.
\newblock


\bibitem[\protect\citeauthoryear{Whittaker}{Whittaker}{2019}]%
        {dashlane-2}
\bibfield{author}{\bibinfo{person}{Zack Whittaker}.}
  \bibinfo{year}{2019}\natexlab{}.
\newblock \bibinfo{title}{{Password manager Dashlane raises \$110M in Series D,
  adds CMO}}.
\newblock
  \bibinfo{howpublished}{{https://techcrunch.com/2019/05/30/dashlane-series-d/}}.
\newblock


\bibitem[\protect\citeauthoryear{Yingtong~Dou and Yu}{Yingtong~Dou and
  Yu}{2019}]%
        {DouDownFraud19}
\bibfield{author}{\bibinfo{person}{Zhirong Liu Zhenhua Dong Jiebo~Luo
  Yingtong~Dou, Weijian~Li} {and} \bibinfo{person}{Philip~S. Yu}.}
  \bibinfo{year}{2019}\natexlab{}.
\newblock \showarticletitle{{Uncovering download fraud activities in mobile app
  markets}}. In \bibinfo{booktitle}{\emph{IEEE/ACM International Conference on
  Advances in Social Networks Analysis and Mining}}.
\newblock


\bibitem[\protect\citeauthoryear{Zhu, Xiong, Ge, and Chen}{Zhu
  et~al\mbox{.}}{2013}]%
        {zhu2013ranking}
\bibfield{author}{\bibinfo{person}{Hengshu Zhu}, \bibinfo{person}{Hui Xiong},
  \bibinfo{person}{Yong Ge}, {and} \bibinfo{person}{Enhong Chen}.}
  \bibinfo{year}{2013}\natexlab{}.
\newblock \showarticletitle{Ranking fraud detection for mobile apps: A holistic
  view}. In \bibinfo{booktitle}{\emph{Proceedings of the 22nd ACM international
  conference on Information \& Knowledge Management}}. ACM,
  \bibinfo{pages}{619--628}.
\newblock


\bibitem[\protect\citeauthoryear{Ziv}{Ziv}{2015}]%
        {incentivize_discussion}
\bibfield{author}{\bibinfo{person}{Tai Ziv}.} \bibinfo{year}{2015}\natexlab{}.
\newblock \bibinfo{title}{{What’s Better? Incentivized Or Non-Incentivized
  App-Install Campaigns}}.
\newblock
  \bibinfo{howpublished}{{https://techcrunch.com/2015/09/10/whats-better-incentivized-or-non-incentivized-app-install-campaigns/}}.
\newblock


\bibitem[\protect\citeauthoryear{{Ziv Bass Specktor}}{{Ziv Bass
  Specktor}}{2020}]%
        {thirdpartymediator}
\bibfield{author}{\bibinfo{person}{{Ziv Bass Specktor}}.}
  \bibinfo{year}{2020}\natexlab{}.
\newblock \bibinfo{title}{{Attribution model explained}}.
\newblock
  \bibinfo{howpublished}{{\url{https://support.appsflyer.com/hc/en-us/articles/207447053-Attribution-model-explained}}}.
\newblock


\end{thebibliography}



\end{document}